\documentstyle [twocolumn,prb,aps,graphicx,floats] {revtex}

\begin{document}
\draft
\begin{title}
{Group theoretical analysis of symmetry breaking in two-dimensional 
quantum dots}
\end{title} 
\author{Constantine Yannouleas and Uzi Landman} 
\address{
School of Physics, Georgia Institute of Technology,
Atlanta, Georgia 30332-0430 }
\date{February 2003; Revised:March 2003}
\maketitle
\begin{abstract}
We present a group theoretical study of the symmetry-broken unrestricted 
Hartree-Fock orbitals and electron densities in the case of a two-dimensional 
$N$-electron single quantum dot (with and without an external magnetic field). 
The breaking of rotational symmetry results in canonical orbitals that (1) are
associated with the eigenvectors of a H\"{u}ckel hamiltonian having sites at 
the positions determined by the equilibrium molecular configuration of the 
classical $N$-electron problem, and (2) transform according to the 
irreducible representations of the point group specified by the discrete 
symmetries of this classical molecular configuration. Through restoration of 
the total-spin and rotational symmetries via post-Hartree-Fock projection 
techniques, we show that the point-group discrete symmetry of the unrestricted
Hartree-Fock wave function underlies the appearance of magic angular momenta 
(familiar from exact-diagonalization studies) in the excitation spectra of the
quantum dot. Furthermore, this two-step symmetry-breaking/symmetry-restoration 
method accurately describes the energy spectra associated with the magic
angular momenta.
\end{abstract}
~~~~\\
\narrowtext

\section{Introduction}

\subsection{Background on the mean-field breaking of spatial symmetries in 
quantum dots}

Two-dimensional (2D) quantum dots (QD's) created at semiconductor interfaces
with refined control of their size, shape and number of electrons are often 
referred\cite{kas,ash,tar,kou} to as ``artificial atoms''. For high magnetic 
fields ($B$), it has been known for some time (ever since the pioneering work 
\cite{lau2} of Laughlin in 1983 concerning the fractional quantum Hall effect)
that 2D few-electron systems exhibit complex strongly-correlated many-body
physics. Nevertheless, for low magnetic fields, the term artificial atoms
was used initially to suggest that the physics of electrons in such
man-made nanostructures is exclusively related to that underlying the
traditional description\cite{cs} of natural atoms (pertaining particularly
to electronic shells and the Aufbau principle), where the electrons are taken
\cite{hart} to be moving in a spherically averaged effective central mean
field. This traditional picture was given support by experimental 
studies\cite{tar,kou} on vertical QD's, which were followed\cite{thtr} by a 
series of sophisticated theoretical investigations yielding results conforming
to it. 

However, in 1999, the circular (for 2D QD's) 
central-mean-field picture was challenged
by the discovery of solutions with broken {\it space} symmetry in the context 
of spin-and-space unrestricted Hartree-Fock (sS-UHF) mean-field 
calculations.\cite{yl3,yl4} These broken symmetry (BS) solutions appear
spontaneously (due to a phase transition) when the strength of the 
interelectron repulsion relative to the zero-point kinetic energy ($R_W$) 
exceeds a certain critical value. They indicate formation of Wigner 
(or electron) molecules (WM's or EM's) with the electrons located at 
the vertices of nested regular polygons (often referred to as concentric 
rings), familiar from studies of classical point charges.\cite{bp,hg} Such 
molecules were characterized by us as strong or weak Wigner crystallites 
depending on their rigidity or lack thereof (i.e., floppiness), suggesting 
the existence of additional\cite{loz} ``phase transitions'' as a function of 
the parameter $R_W$ (see section II.B for its precise definition). 
Furthermore, it was noted\cite{yl3} that the symmetry breaking should 
be accompanied by the emergence of a spectrum of collective rovibrational 
excitations (finite-size analogs of the Goldstone modes). A subsequent 
investigation\cite{yl7} based on exact solutions for a Helium QD (2e QD) 
confirmed these results and provided a systematic study of the molecular 
rovibrational collective spectra and their transition to independent particle 
excitations, as the rigidity of the WM was reduced through variation of the 
controlling parameter $R_W$.

We remark here that the lower-energy BS UHF solutions already 
capture\cite{yl4} part of the correlation energy, compared to the restricted 
HF (RHF) ones. Improved numerical accuracy has been achieved in subsequent 
studies\cite{yl5,yl6,yl9} through the restoration of the broken symmetries via 
projection techniques. Consequently, the methodology of symmetry breaking at 
the UHF mean-field level and of subsequent symmetry restoration via post 
Hartree-Fock methods\cite{note7} provides a systematic controlled hierarchy of
approximations toward the exact solution, with anticipated advantages for 
complex many electron systems (under field-free conditions\cite{yl5} and in 
the presence of a magnetic field\cite{yl6,yl9}), whose treatment is 
computationally prohibitive with other methods (e.g., exact diagonalization).

\subsection{Background on the mean-field breaking of symmetries in other 
finite-size fermionic systems}

The mean field approach has been a useful tool in elucidating the physics of 
small fermionic systems, from natural atoms and atomic nuclei to metallic 
nanoclusters, and most recently of two-dimensional quantum dots. Of particular
interest for motivating the present work (due to spatial-symmetry-breaking 
aspects) has been the mean-field description of deformed nuclei and 
metal clusters (exhibiting ellipsoidal shapes). At a first level, deformation 
effects in these systems can be investigated via semi-empirical mean-field 
models, like the particle-rotor model\cite{bm1} of Bohr and Mottelson 
(nuclei), the anisotropic-harmonic-oscillator model of Nilsson 
(nuclei\cite{nil} and metal clusters\cite{cle}), and the 
shell-correction method of Strutinsky (nuclei\cite{str} and metal 
clusters\cite{yl1,yl2}). 
At the microscopic level, the mean field is often described\cite{rs,so} 
via the self-consistent single-determinantal Hartree-Fock (HF) theory. At this 
level, however, the description of deformation effects mentioned above 
requires\cite{rs,rs2} consideration of unrestricted Hartree-Fock wave 
functions that break explicitly the rotational symmetries of the original 
many-body hamiltonian, but yield HF determinants with lower energy compared to
the symmetry-adapted restricted Hartree-Fock solutions.

In earlier publications,\cite{yl3,yl4,yl5,yl6,yl9} we have shown that, in the 
strongly correlated regime, UHF solutions violating the rotational (circular) 
symmetry arise most naturally in the case of 2D single QD's 
and for both the cases of zero and high magnetic field.\cite{mk} Unlike the 
case of atomic nuclei, however, where symmetry breaking is associated with 
quadrupole deformations, spontaneous symmetry breaking in 2D QD's induces 
electron localization associated with formation of WM's.

The violation in the mean-field approximation of the symmetries of the 
original many-body hamiltonian  appears to be paradoxical at a first glance.
However, for the specific cases arising in Nuclear Physics and Quantum
Chemistry, two theoretical developments have resolved this 
paradox. They are: (1) the theory of restoration of broken symmetries via 
projection techniques,\cite{py,low} and (2) the group theoretical analysis of 
symmetry-broken HF orbitals and solutions in chemical reactions initiated by 
Fukutome and coworkers,\cite{fuk} who used of course the symmetry groups 
associated with the natural 3D molecules.
Despite the different fields, the general principles established in these 
earlier theoretical developments have provided a wellspring of assistance in 
our investigations of symmetry breaking in QD's. In particular, the 
restoration of broken circular symmetry in the case of single QD's has already
been demonstrated by us in three recent publications.\cite{yl5,yl6,yl9,yl8} 

\subsection{Content of this paper}

In the present paper, we will provide an in-depth group theoretical analysis 
of broken-symmetry UHF orbitals and electron densities (ED's) in the case
of single parabolic QD's. We will show that such an analysis provides further 
support for our earlier interpretation concerning the spontaneous formation of 
collectively rotating electron (or Wigner) molecules (REM's). 
Indeed we will demonstrate deep analogies between the electronic structure of 
the WM and that of the natural 3D molecules. 
In particular, we will show that the breaking of rotational symmetry results 
in canonical UHF orbitals that are associated with the eigenvectors of a 
molecular-type H\"{u}ckel hamiltonian having ``atomic''
sites at positions specified by the equilibrium configuration of the
classical $N$-electron problem; these ``atomic'' sites are localization sites
for the electrons, and they do not imply the presence of a positive nucleus.
Thus, in contrast to the fully delocalized and
symmetry-adapted, independent-particle-model-type orbitals of the RHF,
the BS UHF orbitals with the {\it same spin direction\/} 
resemble closely the molecular orbitals (MO's) formed by linear combinations 
of atomic orbitals (LCAO's), which are prevalent\cite{cot} in Chemistry. 
(Naturally, the LCAO behavior of the UHF orbitals with the same spin direction
allows for a more precise understanding of the term ``electron localization'' 
used by us in previous publications.) An important conclusion of the 
present paper is that the BS UHF orbitals are not necessarily unique; what 
matters, in analogy with the LCAO-MO's of natural molecules, is that they 
transform according to the irreducible representations of the point group 
specified by the discrete symmetries of the classical equilibrium 
configuration.\cite{bp}

Our study leads to the following two main results: (i) in analogy with 3D
natural molecules, the WM's can rotate and the restoration of the total-spin 
and rotational symmetries via projection techniques describes their lowest 
rotational bands ({\it yrast\/} bands\cite{yl7,note9}) in addition 
to the ground state, and (ii) the lowering of the symmetry, which results in 
the (discrete) point-group symmetry of the UHF wave function, underlies the 
appearance of the sequences of magic angular momenta (familiar from 
exact-diagonalization studies\cite{gir,mc,rua,sek,mak,haw}) in the excitation 
spectra of single QD's. Since exact-diagonalization 
methods are typically restricted to small sizes 
with $N \leq 10$, the present two-step method of breakage and subsequent 
restoration of symmetries offers a promising new avenue for accurately 
describing larger 2D electronic systems. A concrete example of the potential 
of this approach is provided by Ref.\ \onlinecite{yl6} and Ref.\ 
\onlinecite{yl9}, where 
our use of the the symmetry-breaking/symmetry-restoration method yielded 
analytic expressions for correlated wave functions that offer a better 
description of the $N$-electron problem in high magnetic fields compared 
to the Jastrow-Laughlin\cite{lau2} expression.

Since the group-theoretical aspects of symmetry breaking at the mean-field
level (and their relation to the properties of the exact solutions) remain a 
vastly unexplored territory in the area of condensed-matter finite-size
systems, in the following we will present an introductory 
investigation of them through a series of rather
simple, but nontrivial, illustrative examples from the field of 2D parabolic 
QD's. The plan of the paper is as follows: Section II reviews briefly the set 
of UHF equations employed by us; Section III and section IV present the case 
of three electrons in the absence and in the presence of an external magnetic 
field $B$, respectively. The more complicated case of six electrons at $B=0$ 
is investigated in section V, while section VI discusses the companion step
of the restoration of broken symmetries, which underlies the appearance of
magic angular momenta in the exact spectra. Finally, section VII presents
a summary. 

Before leaving the Introduction, we wish to stress that analogs of several of
the group theoretical concepts and manipulations employed in this paper can be
found in textbooks concerning standard applications of symmetry groups to the 
electronic structure and chemistry of 3D natural molecules. Here, however, we 
will use these otherwise well-known group theoretical aspects to elucidate the
molecular interpretation of the BS UHF determinants (and associated orbitals) 
in the unexpected context of a newly arising area of physics, namely the 
physics of strong correlations in 2D circular artificial atoms (QD's).

\section{The UHF equations}

\subsection{The Pople-Nesbet equations}

The UHF equations we are using are the Pople-Nesbet\cite{pn}
equations described in detail in Ch. 3.8 of Ref.\ \onlinecite{so}.
For completeness, we present here a brief description of them, along with
details of their computational implementation by us to the 2D case
of semiconductor QD's.

The key point is that electrons of $\alpha$ (up) spin are described by one set
of spatial orbitals $\{ \psi^\alpha_j| j=1,2,...,K\}$, while electrons of
$\beta$ (down) spin are described by a different set of spatial orbitals
$\{ \psi^\beta_j| j=1,2,...,K\}$ (of course in the RHF $\psi^\alpha_j =
\psi^\beta_j=\psi_j$). Next, one introduces a set of basis functions 
$\{ \varphi_\mu| \mu=1,2,...,K\}$ (constructed to be 
{\it orthonormal\/} in our 2D case), and expands the UHF orbitals as
\begin{equation}
\psi^\alpha_i = \sum_{\mu=1}^K C_{\mu i}^\alpha \varphi_\mu,~~~i=1,2,...,K
\label{expa}
\end{equation}
\begin{equation}
\psi^\beta_i = \sum_{\mu=1}^K C_{\mu i}^\beta \varphi_\mu,~~~i=1,2,...,K.
\label{expb}
\end{equation}

The UHF equations are a system of two coupled matrix eigenvalue problems,
\begin{equation}
{\bf F}^\alpha {\bf C}^\alpha = {\bf C}^\alpha {\bf E}^\alpha 
\label{uhfa}
\end{equation}
\begin{equation}
{\bf F}^\beta {\bf C}^\beta = {\bf C}^\beta {\bf E}^\beta, 
\label{uhfb}
\end{equation}
where ${\bf F}^{\alpha(\beta)}$ are the Fock-operator matrices and 
${\bf C}^{\alpha(\beta)}$ are the vectors formed with the coefficients in the 
expansions (\ref{expa}) and (\ref{expb}). The matrices 
${\bf E}^{\alpha(\beta)}$ are {\it diagonal\/}, and as a result equations
(\ref{uhfa}) and (\ref{uhfb}) are {\it canonical\/} (standard). Notice that
noncanonical forms of HF equations are also possible (see Ch. 3.2.2 of Ref.\
\onlinecite{so}). Since the selfconsistent iterative solution of the HF
equations can be computationally implemented only in their canonical form,
heretofore canonical orbitals and solutions will always be implied, unless 
otherwise noted explicitly. We note that the coupling between the two UHF
equations (\ref{uhfa}) and (\ref{uhfb}) is given explicitly in the
expressions for the elements of the Fock matrices below [Eqs.\ (\ref{famn})
and (\ref{fbmn})].

Introducing the density matrices ${\bf P}^{\alpha(\beta)}$ for $\alpha(\beta)$
electrons,
\begin{equation}
P^\alpha_{\mu\nu} = \sum_{a}^{N^\alpha}
C^\alpha_{\mu a} (C^\alpha_{\nu a})^*
\label{ppa}
\end{equation}
\begin{equation}
P^\beta_{\mu\nu} = \sum_{a}^{N^\beta} 
C^\beta_{\mu a} (C^\beta_{\nu a})^*,
\label{ppb}
\end{equation}
where $N^\alpha + N^\beta = N$,
the elements of the Fock-operator matrices are given by
\begin{eqnarray}
F^\alpha_{\mu \nu} = H_{\mu \nu} &+& 
\sum_\lambda \sum_\sigma P^\alpha_{\lambda \sigma}
[(\mu \sigma | \nu \lambda ) - (\mu \sigma | \lambda \nu)] \nonumber \\
&+& \sum_\lambda \sum_\sigma P^\beta_{\lambda \sigma}
(\mu \sigma | \nu \lambda ) 
\label{famn}
\end{eqnarray}
\begin{eqnarray}
F^\beta_{\mu \nu} = H_{\mu \nu} &+&
\sum_\lambda \sum_\sigma P^\beta_{\lambda \sigma}
[(\mu \sigma | \nu \lambda ) - (\mu \sigma | \lambda \nu)] \nonumber \\
&+& \sum_\lambda \sum_\sigma P^\alpha_{\lambda \sigma}
(\mu \sigma | \nu \lambda ),
\label{fbmn}
\end{eqnarray}
where $H_{\mu \nu}$ are the elements of the single electron hamiltonian
with an external magnetic field $B$ and an appropriate potential confinement,
and the Coulomb repulsion is expressed via the two-electron integrals
\begin{eqnarray}
(\mu \sigma && | \nu \lambda) = \nonumber \\
&& \frac{e^2}{\kappa}
\int d{\bf r}_1 d{\bf r}_2 \varphi^*_\mu({\bf r}_1) \varphi^*_\sigma({\bf r}_2)
\frac{1}{|{\bf r}_1 - {\bf r}_2|} 
\varphi_\nu({\bf r}_1) \varphi_\lambda({\bf r}_2),
\label{r12}
\end{eqnarray}
with $\kappa$ being the dielectric constant of the semiconductor material.
Of course, the Greek indices $\mu$, $\nu$, $\lambda$, and $\sigma$ run from
1 to $K$. 

For a QD, the external confinement is assumed to be parabolic, and the 
single-particle hamiltonian in a perpendicular  external magnetic field $B$ is
written as 
\begin{equation}
H=\frac{({\bf p}-e{\bf A}/c)^2}{2m^*} + 
\frac{1}{2} m^* \omega^2_{0} (x^2 + y^2) + 
\frac{g^* \mu_B}{\hbar} {\bf B \cdot s}.
\label{hsp}
\end{equation}
The vector potential ${\bf A}$ is given in the symmetric gauge by 
\begin{equation}
{\bf A}({\bf r})=\frac{1}{2}{\bf B} \times {\bf r} =\frac{1}{2}(-By,Bx,0), 
\label{vectp}
\end{equation}
and the last term in Eq. (\ref{hsp}) is the Zeeman interaction with $g^*$ 
being the effective $g$ factor, $\mu_B$ the Bohr magneton, and ${\bf s}$ the 
spin of an individual electron. $m^*$ is the effective electron mass and 
$\omega_0$ is the frequency parameter of the parabolic potential confinement.
 
The system of the two coupled UHF matrix equations 
(\ref{uhfa}) and (\ref{uhfb})
is solved selfconsistently through iteration cycles.\cite{note11}
For obtaining the numerical solutions, we have 
used a set of $K=78$ basis states $\varphi_i$'s 
that are chosen to be the product wave functions formed out from the
eigenstates of the one-dimensional harmonic oscillators along the
$x$ and $y$ axes. Note that the value $K=78$ corresponds to all the 
states of the associated 2D harmonic oscillator up to and including the 12th 
major shell.

\begin{figure}[t]
\centering\includegraphics[width=8.5cm]{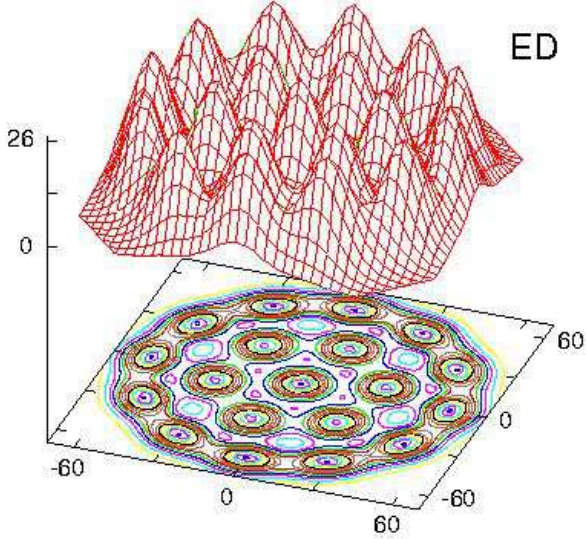}\\
~~~~\\
\caption{
UHF electron density in a parabolic QD for $N=19$ and $S_z=19/2$, exhibiting 
breaking of the circular symmetry at $R_W=5$ and $B=0$. 
The choice of the remaining parameters is: $\hbar \omega_0=5$ meV 
and $m^*=0.067 m_e$.
Distances are in nanometers and the electron density in $10^{-4}$ nm$^{-2}$.
}
\end{figure}
Having obtained the selfconsistent solution, the total UHF energy is
calculated as
\begin{equation}
E_{\text{UHF}}= \frac{1}{2} \sum_\mu \sum_\nu
[ (P^\alpha_{\nu\mu}+P^\beta_{\nu\mu}) H_{\mu\nu}
   + P^\alpha_{\nu\mu} F^\alpha_{\mu\nu}
   + P^\beta_{\nu\mu} F^\beta_{\mu\nu}].
\label{euhf}
\end{equation}

\subsection{Solutions representing Wigner molecules}

As a typical example of a solution that can be extracted from the above 
UHF equations, we mention the case of $N=19$ electrons for 
$\hbar \omega_0 = 5$ meV, $R_W=5$ $(\kappa=3.8191)$, and $B=0$. The Wigner 
parameter $R_W \equiv Q/\hbar \omega_0$, where $Q$ is the Coulomb interaction 
strength; $Q=e^2/\kappa l_0$, with $l_0=(\hbar/m^* \omega_0)^{1/2}$ being the 
spatial extent of the lowest single-electron wave function in the parabolic 
confinement.

Fig.\ 1 displays the total electron density of the BS UHF solution for these 
parameters, which exhibits breaking of the rotational symmetry. In accordance 
with ED's for smaller dot sizes published by us earlier,\cite{yl3,yl4} 
the ED in Fig.\ 1 is highly suggestive of the formation of a Wigner 
molecule, with an (1,6,12) ring structure in the present case. 
This polygonal ring structure agrees with the 
classical one\cite{bp} and is sufficiently complex to instill confidence that 
the Wigner-molecule interpretation is valid. The following question, however,
arises naturally at this point: is such molecular interpretation limited to 
the intuition provided by the landscapes of the total ED's, or are there 
deeper analogies with the electronic structure of natural 3D molecules? The 
answer to the second part of this question is in the positive, and the 
remainder of this paper is devoted to discovering such analogies. However, 
since the $N=19$ case represents a rather complicated group-theoretical 
structure, for simplicity and transparency, we will study in the following 
smaller QD sizes. This, however, will not result in any loss of generality in 
our conclusions. 

In previous publications,\cite{yl3,yl4} we found that {\it space\/} symmetry 
breaking in the UHF equations appears spontaneously 
for $R_W > 1$. Below we choose to work 
with larger values of $R_W$ (e.g., 10 or higher), for which the effects of
strong correlations are fully developed. The group-theoretical investigation 
of the intermediate regime near the phase transition is left for a future 
publication. In all calculated cases, we used $\hbar \omega_0=5$ meV,  
$m^*=0.067 m_e$ (GaAs), and $g^*=-0.44$ (GaAs).

We note that the Pople-Nesbet UHF equations
are primarily employed in Quantum Chemistry for studying the ground states of
open-shell molecules and atoms. Unlike our studies of QD's, however, such
chemical UHF studies consider mainly the breaking of the total spin 
symmetry, and not that of the space symmetries. As a result, for purposes of 
emphasis and clarity, we have often used (see, e.g., our previous papers) 
prefixes to indicate the specific unrestrictions involved in our UHF 
solutions, i.e., the prefix s- for the total-spin and the prefix S- for the 
space unrestriction. 

\begin{figure}[t]
\centering\includegraphics[width=7.8cm]{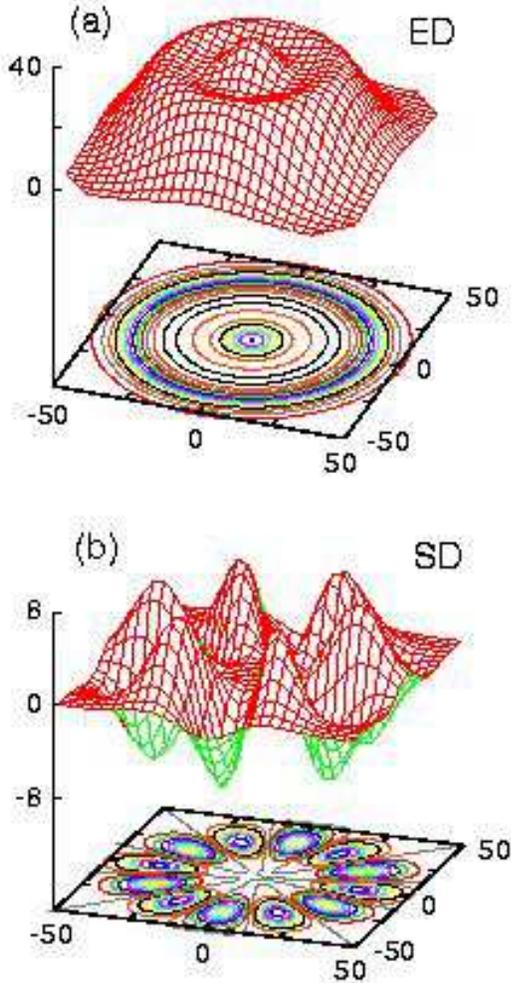}\\
~~~~\\
\caption{
UHF solution in a parabolic QD exhibiting a pure spin density wave for $N=14$,
$S_z=0$, $R_W=0.8$, and $B=0$. (a) The total electron density exhibiting
circular symmetry; (b) The spin density exhibiting azimuthal modulation
(note the 12 humps whose number is smaller than the number of electrons).
The choice of the remaining parameters is: $\hbar \omega_0=5$ meV 
and $m^*=0.067 m_e$.
Distances are in nanometers and the electron and spin densities in 
$10^{-4}$ nm$^{-2}$.
}
\end{figure}

\subsection{Solutions representing pure spin density waves}

\begin{figure}[t]
\centering\includegraphics[width=8.5cm]{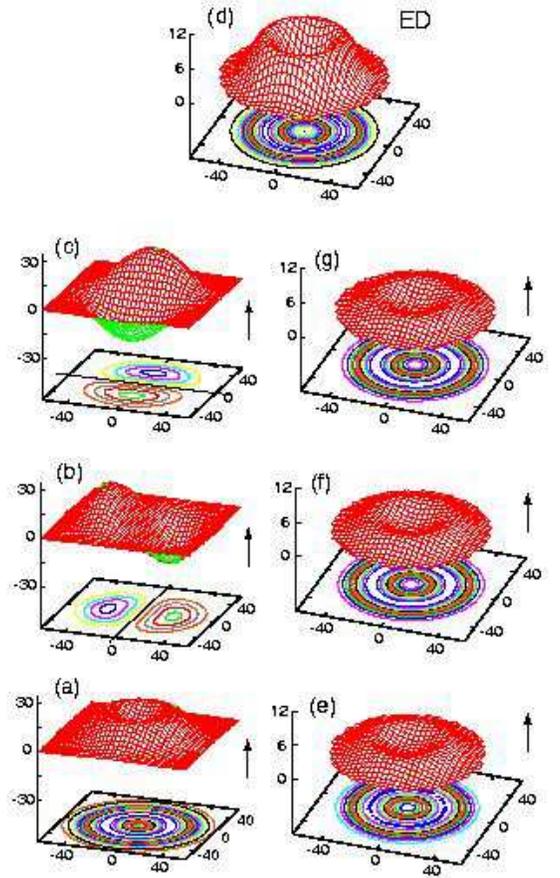}\\
~~~~\\
\caption{
The RHF case for $N=3$ and $S_z=3/2$ at $R_W=10$. (a-c): real orbitals at 
$B=0$. (e-g): the modulus square of the complex orbitals for a vanishingly 
small value $B=0.0001$ T. (d): the corresponding circular electron density
for both cases. 
The choice of the remaining parameters is: $\hbar \omega_0=5$ meV 
and $m^*=0.067 m_e$.
Distances are in nanometers. The real orbitals are in
10$^{-3}$ nm$^{-1}$ and the orbital densities and total ED in 
10$^{-4}$ nm$^{-2}$. 
The arrows indicate the spin direction.
}
\end{figure}
Before leaving this section, we mention another class of BS solutions, which
can appear in single QD's, namely the spin density waves (SDW's). The SDW's 
are unrelated to electron localization and thus are quite distinct from the 
WM's; in single QD's, they were obtained\cite{kmr} earlier within the 
framework of spin density functional theory.\cite{note4} To emphasize the 
different nature of SDW's and WM's, we present in Fig.\ 2 an example of a
SDW obtained with the UHF approach [the corresponding parameters are: $N=14$, 
$S_z=0$, $R_W=0.8$ ($\kappa=23.8693$), and $B=0$]. Unlike the case of WM's, 
the SDW exhibits a circular ED [see Fig.\ 2(a)], and thus does not break the 
rotational symmetry. Naturally, in keeping with its name, the SDW breaks the 
total spin symmetry and exhibits azimuthal modulations in the spin 
density\cite{over} (SD) [see Fig.\ 2(b); however, the number of humps is 
smaller than the number of electrons\cite{note3}]. The SDW's in single QD's 
appear for $R_W \lesssim 1$ and are of lesser importance;\cite{note5} thus in 
the following we will exclusively study the case of WM's.

\section{Three electrons without magnetic field}

\subsection{The $S_z=3/2$ fully spin polarized case}

We begin with the case of $N=3$ fully spin polarized $(S_z=3/2)$ electrons
in the absence of a magnetic field and for $R_W=10$ $(\kappa=1.9095)$. Fully
spin polarized UHF determinants preserve the total spin, but for this
value of $R_W$ the lowest in energy UHF solution is one with broken circular 
symmetry. As it will be seen below, broken rotational symmetry does not imply 
no space symmetry, but a lower point-group symmetry. Before proceeding with 
the study of the BS solution, however, 
it will be helpful to review the symmetry-adapted RHF solution first. This RHF
solution can be obtained from the UHF equations (\ref{uhfa}) and (\ref{uhfb}) 
by using a circular electron-density guess as the input of the first iteration.
In the independent particle model, the $N=3$, $S_z=3/2$, and $B=0$ 2D case 
corresponds to a closed electronic shell with configuration $1s1p_+1p_-$ or 
$1s1p_x1p_y$ $(p_\pm \propto p_x \pm i p_y \propto r e^{\pm i\theta})$, and 
thus the independent-particle-model ED is 
necessarily circular. We will confirm that the RHF solution conforms indeed to
the prediction of the independent particle model, and subsequently we will 
contrast the UHF solution to it.

\begin{figure}[t]
\centering\includegraphics[width=8.5cm]{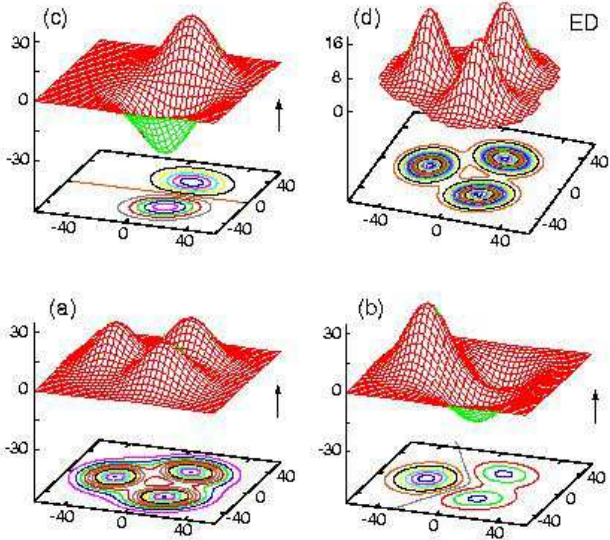}\\
~~~~\\
\caption{
The S-UHF case exhibiting breaking of the circular symmetry for $N=3$ and 
$S_z=3/2$ at $R_W=10$ and $B=0$. (a-c): real orbitals. (d): the corresponding 
electron density.
The choice of the remaining parameters is: $\hbar \omega_0=5$ meV 
and $m^*=0.067 m_e$.
Distances are in nanometers. The real orbitals are in
10$^{-3}$ nm$^{-1}$ and the total ED in 10$^{-4}$ nm$^{-2}$.
The arrows indicate the spin direction.
}
\end{figure}

The RHF solution for $R_W=10$ and $B=0$ has an energy of 
92.217 meV; the corresponding orbitals are real and are displayed in the left 
column of Fig.\ 3. They are like the $1s$ [Fig.\ 3(a)], $1p_x$ [Fig.\ 3(b)], 
and $1p_y$ [Fig.\ 3(c)] orbitals of the independent particle model. 
The nodeless $1s$ orbital has a 
maximum at $r > 0$ due to the large Coulomb repulsion; its energy is 44.526 
meV. The energy of the two degenerate $p_x$ and $p_y$ orbitals is 50.489 meV.
Notice that neither the $p_x$ nor $p_y$ orbital is rotationally symmetric;
however, the total ED [Fig.\ 3(d)] has the expected circular symmetry. It is 
of interest to obtain the RHF solution for a very small external magnetic 
field (i.e., in the limit $B \rightarrow 0$). In this case, the calculated 
total and orbital energies, as well as the total ED [Fig.\ 3(d)] remain 
unchanged. However, the two degenerate $p$ orbitals are now complex [$p_\pm$, 
see Fig.\ 3(f) and Fig.\ 3(g)] and have good angular momenta $l=\pm 1$, and 
thus their modulus square is circularly symmetric.

We focus now on the sS-UHF solution for $N=3$ and $S_z=3/2$, and for the 
same choice of parameters as with the RHF case. The UHF total  
energy is 89.691 meV, and thus it is lower than the corresponding RHF one.
In Fig.\ 4 we display the UHF symmetry-violating orbitals (a$-$c) whose 
energies are (a) 44.801 meV, (b) and (c) 46.546 meV, namely the two orbitals 
(b) and (c) with the higher energies are again degenerate in energy. 
An inspection of Fig.\ 4 immediately reveals that these orbitals have retained
some properties of the delocalized $1s$, $1p_x$ and $1p_y$ orbitals, familiar 
from the independent particle model and the RHF; that is, orbital (a) is 
nodeless, while each one of the orbitals (b) and (c) has a single nodal line.
However, overall the BS orbitals (a$-$c) drastically differ from the orbitals 
of the independent particle model. 
In particular, they are associated with specific sites (within the
QD) forming an equilateral triangle, and thus they can be described as having 
the structure of a linear combination of ``atomic'' (site) orbitals (LCAO's). 
Such LCAO molecular orbitals (MO's) are familiar in natural molecules, and this
analogy supports the term ``electron (or Wigner) molecules'' for 
characterizing the BS UHF solutions. 

In the LCAO-MO approximation, one needs to solve a matrix eigenvalue equation
determined by the overlaps $S_{ij}$ ($i \neq j$) and the hamiltonian matrix 
elements $\widetilde{H}_{ij}$ and $\widetilde{H}_{ii}$  
between the atomic orbitals. A further simplified approximation,\cite{note1} 
the H\"{u}ckel approximation (H\"{u}A), consists in taking all $S_{ij}=0$, 
and all $\widetilde{H}_{ij}=0$ unless the $i$th and $j$th atoms 
(sites) are adjacent. For our example this latter approximation is applicable,
since the value $R_W=10$ is rather high. When using the notations
$\epsilon= \widetilde{H}_{11}=\widetilde{H}_{22}=\widetilde{H}_{33}$ and
$-\beta= \widetilde{H}_{12}=\widetilde{H}_{13}=\widetilde{H}_{23} < 0$, 
the H\"{u}ckel eigenvalue equation for the case of $N=3$ electrons on the 
vertices of an equilateral triangle is written as 
\begin{equation}
\left( \begin{array}{ccc}
         \epsilon & -\beta      & -\beta \\
           -\beta & \epsilon    & -\beta \\
           -\beta & -\beta      & \epsilon  
       \end{array}   \right)
\left( \begin{array}{c}
          f_1 \\ f_2 \\ f_3
        \end{array}   \right)
= E \left( \begin{array}{c}
              f_1 \\ f_2 \\ f_3
           \end{array}   \right),
\label{meqn3}
\end{equation}
and the associated LCAO-MO orbitals are $\psi_i = f^i_1 \phi_1 +
f^i_2 \phi_2 + f^i_3 \phi_3$, having $E_i$ eigenvalues with
$i=1,2,3$. The $\phi_j$'s are the original Gaussian-type atomic (site) 
orbitals.

From the eigenvalues and eigenvectors of (\ref{meqn3}), one finds the 
following three LCAO-MO's:  
\begin{equation}
\psi_1 = (\phi_1+\phi_2+\phi_3)/\sqrt{3}
\label{psi1}
\end{equation}
with energy $E_1 = \epsilon - 2 \beta$, 
\begin{equation}
\psi_2 =(2\phi_1-\phi_2-\phi_3) /\sqrt{6}
\label{psi2}
\end{equation}
with energy $E_2=\epsilon + \beta$, and
\begin{equation}
\psi_3=(\phi_2-\phi_3)/\sqrt{2}
\label{psi3}
\end{equation}
with energy $E_3=E_2$. It is apparent that the structure of these 
three LCAO-MO's and the level diagram of their energies agrees 
very well with the corresponding symmetry-broken UHF orbitals [displayed in 
Figs.\ 4(a) $-$ 4(c)] and their energies. (Using the HF values for $E_1$ and 
$E_2$, one finds $\epsilon \approx 45.961$ meV and $\beta \approx 0.585$ meV.)
We notice here that such LCAO orbitals are familiar in Organic Chemistry and 
are associated with the theoretical description of Carbocyclic Systems, and in
particular the molecule C$_3$H$_3$ (cyclopropenyl, see, e.g., Ref.\ 
\onlinecite{cot}).

Naturally, since the orbitals (b) and (c) are degenerate in energy, they are 
not uniquely defined: any linear combination associated with a unitary 
$2 \times 2$ transformation will produce a pair of different, but equivalent 
(b$^\prime$) and (c$^\prime$) orbitals. The fact that the UHF orbitals in 
Fig.\ 4 have the specific highly symmetrized (see below) form given above is 
the result of an accidental choice of the initial electron-density input in 
the HF iteration. We have checked that any such pair of (b$^\prime$) and 
(c$^\prime$) orbitals leaves the 2D total UHF electron density unchanged. This
suggests that there is an underlying group theoretical structure that
governs the BS UHF orbitals. The important point is not the uniqueness or not
of the 2D sS-UHF orbitals, but the fact that they transform according to the
irreducible representations of specific point groups, leaving both the
sS-UHF determinant and the associated electron densities invariant.
Given the importance of this observation, we proceed in the rest of this 
section with a group theoretical analysis of the BS sS-UHF orbitals for the 
$N=3$ and $S_z=3/2$ case.

The ED portrayed in Fig.\ 4(d) remains invariant under certain geometrical 
symmetry operations, namely those of an unmarked, 
plane and equilateral triangle. They are: (1) The identity $E$; (2) The two 
rotations $C_3$ (rotation by $2\pi/3$) and $C_3^2$ (rotation by $4 \pi/3$);
and (III) The three reflections $\sigma_v^I$, $\sigma_v^{II}$, and 
$\sigma_v^{III}$ through the three vertical planes, one passing through each
vertex of the triangle. These symmetry operations for the unmarked equilateral
triangle constitute the elements of the group $C_{3v}$.\cite{cot,wol}
 
One of the main applications of group theory in Chemistry is the determination
of the eigenfunctions of the Schr\"{o}dinger equation without actually solving
the matrix equation (\ref{meqn3}). This is achieved by constructing the 
so-called {\it symmetry-adapted linear combinations\/} (SALC's) of AO's. 
A widely used tool for constructing SALC's is the projection operator
\begin{equation}
\hat{{\cal P}}^\mu = \frac{n_\mu}{|{\cal G}|} \sum_R \chi^\mu(R) \hat{R},
\label{prch}
\end{equation}
where $\hat{R}$ stands for any one of the symmetry operations of the molecule,
and $\chi^{\mu}(R)$ are the characters of the 
$\mu$th irreducible representation
of the set of $\hat{R}$'s. (The $\chi^\mu$'s are tabulated in the socalled
character tables.\cite{cot,wol}) $|{\cal G}|$ denotes the order of the group
and $n_\mu$ the dimension of the representation.

The task of finding the SALC's for a set of three $1s$-type AO's exhibiting 
the $C_{3v}$ symmetry of an equilateral triangle can be simplified, since the 
pure rotational symmetry by itself (the rotations $C_3$ 
and $C_3^2$, and not the reflections $\sigma_v$'s through the 
vertical planes) is sufficient for their determination. Thus one needs to 
consider the simpler character table\cite{note6} of the cyclic group $C_3$ 
(see Table I)
\begin{table}[t]
\caption{Character table for the cyclic group $C_3$ [$\varepsilon = 
\exp(2\pi i/3)]$}
\begin{tabular}{c|ccc}
$C_3$ & $E$ & $C_3$ & $C_3^2$ \\ \hline
~~$A$~~ & 1 & 1 & 1 \\
~~$E^\prime$~~ & 1 & $\varepsilon$ & $\varepsilon^*$ \\
~~$E^{\prime\prime}$~~ & 1 & $\varepsilon^*$ & $\varepsilon$ \\
\end{tabular}
\end{table}

From Table I, one sees that the set of the three $1s$ AO's situated 
at the vertices of an equilateral triangle spans the two irreducible
representations $A$ and $E$, the latter one consisting of two associted
one-dimensional representations. To construct the SALC's, one simply
applies the three projection operators $\hat{{\cal P}}^A$, 
$\hat{{\cal P}}^{E^\prime}$, and $\hat{{\cal P}}^{E^{\prime\prime}}$ to
one of the original AO's, let's say the $\phi_1$,
\begin{eqnarray}
\hat{{\cal P}}^A \phi_1 & \approx &
(1) \hat{E} \phi_1 + (1) \hat{C}_3 \phi_1 + (1) \hat{C}_3^2 \phi_1 \nonumber \\
& = & (1) \phi_1 + (1) \phi_2 + (1) \phi_3 \nonumber \\
& = & \phi_1 + \phi_2 + \phi_3,
\label{paf1}
\end{eqnarray}
\begin{eqnarray}
\hat{{\cal P}}^{E^\prime} \phi_1 & \approx &
(1) \hat{E} \phi_1 + (\varepsilon) \hat{C}_3 \phi_1 + 
(\varepsilon^*) \hat{C}_3^2 \phi_1 \nonumber \\
& = & \phi_1 + \varepsilon \phi_2 + \varepsilon^* \phi_3,
\label{paf2}
\end{eqnarray}
\begin{eqnarray}
\hat{{\cal P}}^{E^{\prime\prime}} \phi_1 & \approx &
(1) \hat{E} \phi_1 + (\varepsilon^*) \hat{C}_3 \phi_1 + 
(\varepsilon) \hat{C}_3^2 \phi_1 \nonumber \\
& = & \phi_1 + \varepsilon^* \phi_2 + \varepsilon \phi_3.
\label{paf3}
\end{eqnarray}
The $A$ SALC in Eq.\ (\ref{paf1}) has precisely the same form as the $\psi_1$ 
MO  in Eq.\ (\ref{psi1}), which was determined via a solution of the 
H\"{u}ckel equation (\ref{meqn3}). The two $E$ SALC's [Eq.\ (\ref{paf2}) and 
Eq.\ (\ref{paf3})], however, are complex functions and do not coincide with 
the real $\psi_2$ and $\psi_3$ found above [Eq.\ (\ref{psi2}) and Eq.\
(\ref{psi3})]. As we will see in section IV, these complex SALC's agree with 
BS UHF orbitals obtained in the case of an applied magnetic field. 
On the other hand, a set of two real and orthogonal SALC's 
that spans the $E$ representation can be derived fron Eq.\ (\ref{paf2}) and 
Eq.\ (\ref{paf3}) by simply adding and substracting the two complex ones. This
procedure recovers immediately the real $\psi_2$ and $\psi_3$ discussed 
earlier. 

\begin{figure}[t]
\centering\includegraphics[width=8.5cm]{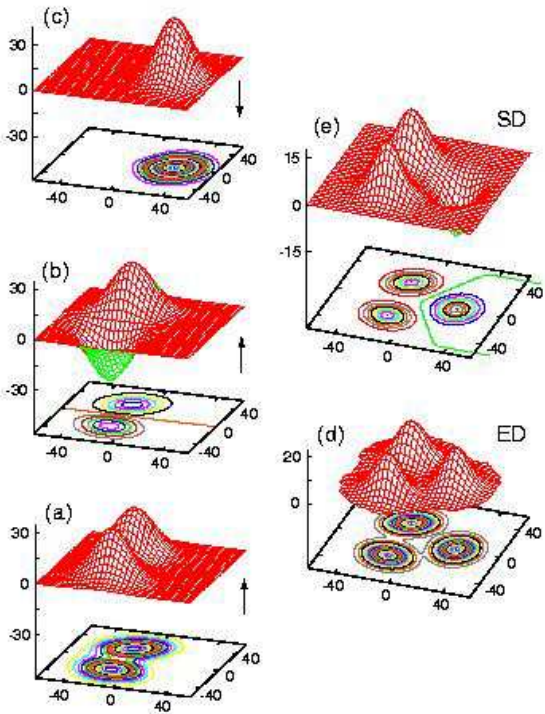}\\
~~~~\\
\caption{
The sS-UHF solution exhibiting breaking of the circular symmetry for $N=3$ and 
$S_z=1/2$ at $R_W=10$ and $B=0$. (a-b): real orbitals for the two 
spin-up electrons. (c): real orbital for the single spin-down electron. 
(d): total electron density. (e): spin density (difference of the spin-up minus
the spin-down partial electron densities).
The choice of the remaining parameters is: $\hbar \omega_0=5$ meV 
and $m^*=0.067 m_e$.
Distances are in nanometers. The real orbitals are in
10$^{-3}$ nm$^{-1}$ and the densities (ED and SD) in 10$^{-4}$ nm$^{-2}$.
The arrows indicate the spin direction.
}
\end{figure}

We stress here that the UHF orbitals of Fig.\ 4 are {\it canonical\/}
(see section II.A). As is well known from Quantum Chemistry,\cite{so}
in general the canonical spin orbitals will be spread out over the different 
sites (atoms) of a natural molecule and will form a basis for the irreducible 
representations of the symmetry group of the molecule. Once the 
canonical orbitals are available, there is an infinite number of {\it 
noncanonical\/} spin orbitals that span {\it reducible\/} representions of the
symmetry group of the molecule and can be obtained via a unitary 
transformation of the canonical set. We remind the reader that noncanonical
spin orbitals are solutions of a generalized HF equation involving 
off-diagonal elements $E_{ij}$ in the matrix formed out of the HF orbital
energies (see section II.A). Naturally, the unitary transformation leaves the 
UHF determinant and total energy invariant. In particular there is a unitary 
matrix that transforms the canonical spin orbitals to the fully localized 
AO's; i.e., for the $N=3$ and $S_z=3/2$ case, such a unitary matrix transforms
the canonical $\psi_i$'s ($i=1,2,3$) to the noncanonical AO's $\phi_i$'s. 

Note that such noncanonical orbitals for $N=3$ were recently 
used\cite{sun} to formulate a non-selfconsistent variant to the Pople-Nesbet 
HF equations listed in section II.A. This variant relied on the manifestation
of spontaneous symmetry breaking that was discovered earlier via our 
selfconsistent UHF results. Notice, however, that Ref.\ \onlinecite{sun} 
obtained an incomplete wave function, since the companion step of restoration 
of the rotational symmetry was not considered (see section VI below).

\subsection{The $S_z=1/2$ partially spin polarized case}

In Fig.\ 5, we display the sS-UHF symmetry-violating orbitals (a$-$c)
for the case of a partially polarized QD at $R_W=10$ and $B=0$ with two 
spin-up and one spin-down electrons ($N=3$ and $S_z = 1/2$). The UHF orbital
energies of these electrons are: (a) 45.350 and (b) 46.515 meV for the spin-up
orbitals, and (c) 45.926 meV for the spin-down orbital. An inspection of
Fig.\ 5 reveals that these orbitals have retained the nodal structure of
the corresponding independent-particle-model 
orbitals in the familiar $1s^2 1p_x$ configuration; 
namely, (a) and (c) are nodeless, while (b) exhibits a single nodal line.
Apart of this property, however, and in consonance with the $S_z=3/2$ case
studied in a previous subsection, the BS UHF orbitals again differ drastically
from the ones associated with the independent particle model; 
again they are associated with three sites within the QD
arranged in an equilateral triangle. In contrast to the fully polarized case,
however, there are no linear combinations of atomic orbitals involving all 
three vertices of the triangle. Indeed the single spin-down electron remains 
by itself as an unmodified AO, while only the two spin-up electrons combine to
form LCAO MO's. This behavior here is a special case of a general property of 
the sS-UHF, i.e., only AO's associated with the same spin direction can
in principle combine to form LCAO MO's. This property, however, does not
extend to the Generalized Hartree-Fock \cite{fuk,and} which 
incorporates the additional unresctriction that the $z$-projection ($S_z$) of 
the total spin is not preserved and it is not a good quantum number.
(Notice that unlike the practice in this paper, Ref.\ \onlinecite{fuk}
uses the term UHF for the Generalized HF.)

In the same spirit with the treatment of the fully polarized case in 
subsection III.A, and taking into consideration the decoupling of the two
different spin directions in the sS-UHF, one can write a corresponding 
H\"{u}ckel matrix equation for the $N=3$ and $S_z=1/2$ case as follows,
\begin{equation}
\left( \begin{array}{ccc}
         \epsilon & -\beta      &  0  \\
           -\beta & \epsilon    &  0  \\
              0   &    0        & \epsilon  
       \end{array}   \right)
\left( \begin{array}{c}
          f_1 \\ f_2 \\ f_3
        \end{array}   \right)
= E \left( \begin{array}{c}
              f_1 \\ f_2 \\ f_3
           \end{array}   \right).
\label{meqn21}
\end{equation}

From the eigenvalues and eigenvectors of (\ref{meqn21}), one finds
the following three LCAO-MO's: (a) $\psi_1 = (\phi_1+\phi_2)/\sqrt{2}$ with
energy $E_1=\epsilon-\beta$, (b) $\psi_2 = (\phi_1-\phi_2)/\sqrt{2}$ with
energy $E_2=\epsilon+\beta$, and (c) $\psi_3=\phi_3$ with energy 
$E_3=\epsilon$. The structure of these three LCAO-MO's agrees very well with 
the corresponding symmetry-violating UHF orbitals for the $N=3$ and $S_z=1/2$ 
case displayed in Figs.\ 5(a) $-$ 5(c). This agreement extends also to the
level diagrams of the corresponding orbital energies.
(Using the UHF values for $E_1$ and $E_2$, one finds $\beta=0.582$
meV and $\epsilon=45.932$ meV $\approx E_3$.)

\begin{figure}[t]
\centering\includegraphics[width=8.5cm]{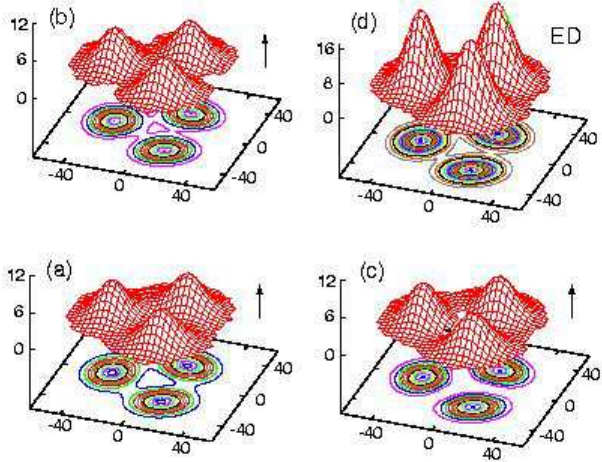}\\
~~~~\\
\caption{
The S-UHF solution exhibiting breaking of the circular symmetry for $N=3$ and 
$S_z=3/2$ at $R_W=10$ and $B=2$ T. (a-c): orbitals (modulus square) for the 
three spin-up electrons. (d): total electron density.
The choice of the remaining parameters is: $\hbar \omega_0=5$ meV, 
$m^*=0.067 m_e$, and $g^*=-0.44$.
Distances are in nanometers and the densities (orbital and ED) in 10$^{-4}$
nm$^{-2}$.
The arrows indicate the spin direction.
}
\end{figure}
Concerning the underlying symmetry-group structure of the UHF
orbitals in Fig.\ 5, we observe that (unlike the fully polarized case) the
two rotations, i.e., $C_3$ (rotation by $2\pi/3$) and $C_3^2$ (rotation by 
$4\pi/3$), are not part of the symmetry operations of the relevant point group
[due to the dissimilarity between spin-up and spin-down orbitals; this can 
be seen clearly through inspection of the spin density in Fig.\ 5(e)].
Because of the 2D character of the dot, the only symmetry operations
for the $N=3$ and $S_z=1/2$ case are the identity $E$ and a single reflection,
$\sigma_v^I$, through the vertical plane passing through the spin-down
electron. Such a group \{$E$, $\sigma_v^I$\} is a subgroup of the familiar 
$C_{2v}$ group associated with the 3D H$_2$O molecule (observe that the 
O atom corresponds to the spin-down electron and the two H atoms correspond to
the two spin-up electrons). According to Ref.\ \onlinecite{cot} (see p. 181), 
the representation $\Gamma$ formed by the three original AO's can be reduced to
irreducible $A_2$ and $B_1$ representations as $\Gamma = A_2 + 2B_1$. By 
applying projection operators [see Eq.(\ref{prch})] to the $\phi_1$ AO and 
using the character Table II, 
\begin{table}[b]
\caption{Character table for the group \{$E$, $\sigma_v^I$\}}
\begin{tabular}{c|cc}
~~~   & $E$   & $\sigma_v^I$  \\ \hline
~~$A_2$~~ & 1 & $-1$ ~~ \\
~~$B_1$~~ & 1 & 1  ~~
\label{tb2}
\end{tabular}
\end{table}
one finds the following two normalized SALC's,
\begin{equation} 
\psi_{A_2} = (\phi_1-\phi_2)/\sqrt{2},
\label{pa2}
\end{equation}
and
\begin{equation} 
\psi_{B_1} = (\phi_1+\phi_2)/\sqrt{2}.
\label{pb11}
\end{equation}
Naturally the second SALC of $B_1$ symmetry is
\begin{equation} 
\psi^{\prime}_{B_1} = \phi_3.
\label{pb12}
\end{equation}

Once more, we stress the fact that the SALC's [Eqs.\ (\ref{pa2})$-$
(\ref{pb12})] derived above via symmetry-group theory have the same structure 
as the canonical UHF orbitals displayed in Fig.\ 5. Observe further that in the
Generalized HF (as also in the case of the H$_2$O molecule and the allyl
anion) the two SALC's of $B_1$ symmetry are allowed to couple, producing more 
complicated orbitals and energy level diagrams
(see p. 1037 of Ref.\ \onlinecite{fuk} and p. 183 in Ref.\ \onlinecite{cot})

\section{Three electrons in a finite magnetic field}

\subsection{The $S_z=3/2$ fully polarized case}

In this section, we study the case of three fully polarized electrons
under a magnetic field $B$. Since for $B \neq 0$, the BS UHF orbitals are
necessarily complex functions, Fig.\ 6 displays the modulus square
of these orbitals. The UHF total ED displayed in Fig.\ 6(d) and the modulus
square of the orbitals exhibit an apparent $C_{3v}$ symmetry as was the case 
at $B=0$ (section III.A). However, {\it the phases of the complex orbitals\/} 
at $B \neq 0$ contribute to a modification of the symmetry group. This 
modification has been studied earlier for the case of infinite crystalline 
systems with periodic space lattices where the electrons occupy Bloch orbitals,
and such studies have led to the consideration of two physically equivalent
group structures, namely the {\it ray groups\/} \cite{bro} and the 
{\it magnetic translation groups\/}. \cite{zak} In our case of a finite
periodic crystallite, corresponding magnetic rotation groups would be
straightforward to consider. However, in order to appreciate the modifications
introduced by the magnetic field, it will be simpler to modify the H\"{u}ckel 
(tight-binding) hamiltonian according to the Harper-Peierls prescription, 
\cite{pei,har} which accounts for the magnetic gauge transformation when
moving from one crystalline site to another. 

\begin{figure}[t]
\centering\includegraphics[width=8.5cm]{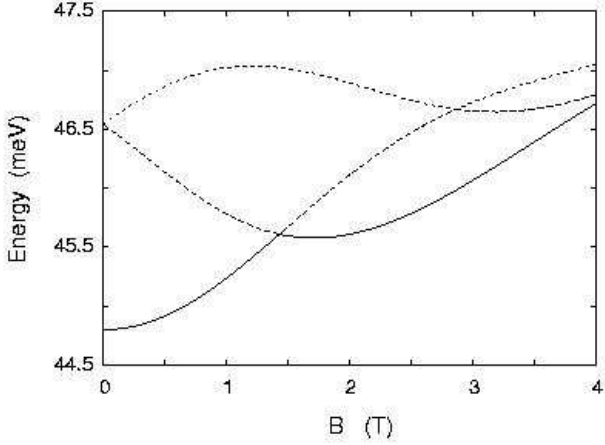}\\
~~~~\\
\caption{
The S-UHF orbital energies (in meV) for $N=3$,$S_z=3/2$ and $R_W=10$ as a 
function of the magnetic field $B$ (in Tesla), exhibiting a prominent 
Aharonov-Bohm oscillation. The choice of the remaining parameters is: 
$\hbar \omega_0 =5$ meV, $m^*=0.067 m_e$, and $g^*=-0.44$.
}
\end{figure}
Thus according to Peierls and
Harper, the proper atomic orbitals $\widetilde{\phi}_j$'s for $B \neq 0$ 
(centered at ${\bf R}_j$) are the real $\phi_j$'s multiplied by an
appropriate phase as follows,
\begin{equation}
\widetilde{\phi}_j ({\bf r}; {\bf R}_j) 
= \phi_j({\bf r}; {\bf R}_j) 
e^{\frac{ie}{\hbar c} {\bf A}({\bf R}_j) \mbox{\boldmath $\cdot$} {\bf r}}.
\label{phase}
\end{equation}
Because of this position-dependent phase in the AO's, the hopping matrix 
elements $\widetilde{H}_{ij}$ (see section III.A) are now complex, and the 
H\"{u}ckel equation (\ref{meqn3}) for three electrons is modified as follows:
\begin{equation}
\left( \begin{array}{ccc}
             \epsilon  & -\beta e^{i\Omega}     & -\beta e^{-i\Omega} \\
  -\beta e^{-i\Omega}  & \epsilon               & -\beta e^{i\Omega} \\
  -\beta e^{i\Omega}  & -\beta e^{-i\Omega}    & \epsilon  
       \end{array}   \right)
\left( \begin{array}{c}
          f_1 \\ f_2 \\ f_3
        \end{array}   \right)
= E \left( \begin{array}{c}
              f_1 \\ f_2 \\ f_3
           \end{array}   \right),
\label{meqn3m}
\end{equation}
where $\Omega \equiv \Omega_{ij} =
(e/\hbar c) {\bf A}({\bf R}_j - {\bf R}_i) 
\mbox{\boldmath $\cdot$} {\bf R}_i$, 
$(i,j)=(1,2),(2,3),(3,1)$, with ${\bf R}_k$, $k=1,2,3$ being the positions of 
the vertices of the equilateral triangle. Notice that $\Omega = (2 \pi \Phi)/
(3 \Phi_0)$, where $\Phi$ is the total magnetic flux through the equilateral 
triangle and $\Phi_0 = hc/e$ is the unit flux.

From the eigenvectors of equation (\ref{meqn3m}), one finds the following 
LCAO-MO's,
\begin{equation}
\psi_1 \propto \widetilde{\phi}_1 + \widetilde{\phi}_2 + \widetilde{\phi}_3,
\label{psi1m}
\end{equation}
\begin{equation}
\psi_2 \propto  e^{2\pi i/3} \widetilde{\phi}_1 
+ e^{-2 \pi i/3} \widetilde{\phi}_2 + \widetilde{\phi}_3,
\label{psi2m}
\end{equation}
\begin{equation}
\psi_3 \propto  e^{-2\pi i/3} \widetilde{\phi}_1 
+ e^{2 \pi i/3} \widetilde{\phi}_2 + \widetilde{\phi}_3,
\label{psi3m}
\end{equation}
with corresponding orbital energies,
\begin{equation}
E_1 = \epsilon - 2 \beta \cos\Omega,
\label{e1m}
\end{equation}
\begin{equation}
E_2 = \epsilon - 2 \beta \cos[(2\pi/3+\Omega)],
\label{e2m}
\end{equation}
\begin{equation}
E_3 = \epsilon - 2 \beta \cos[(2\pi/3-\Omega)].
\label{e3m}
\end{equation}

Substituting the specific value for $\Omega$ given above, one can write 
the eigenvalues (\ref{e1m}-\ref{e3m}) in a more symmetric compact form,
\begin{equation}
E_j= 
\epsilon -2 \beta \cos \left[ \frac{2 \pi}{3} (j+ \frac{\Phi}{\Phi_0}) \right],
\; j=1,2,3.
\label{ejm}
\end{equation}

Since the original AO's do not practically overlap for $R_W=10$, the phases
in front of the $\phi_i$'s in Eqs.\ (\ref{psi1m}-\ref{psi3m}) do not
contribute substantially to the modulus square of the orbitals. As a result,
for all values of $B$, all three orbitals exhibit similar orbital densities 
that are approximately equal to $\phi_1^2 + \phi_2^2 + \phi_3^2$. 
Observe that this agrees very well with the behavior of the canonical UHF 
orbitals (modulus square) at $B=2$ T displayed in Fig.\ 6. At zero magnetic 
field $B=0$, the LCAO-MO's in equations (\ref{psi1m}-\ref{psi3m}) reduce to
the specific form given earlier in Eqs.\ (\ref{paf1}-\ref{paf3}) of section 
III.A. We stress here that in section III.A these LCAO-MO's were derived from 
arguments based exclusively on the group theoretical structure of the $C_{3v}$
symmetry group.

Naturally, when $\Omega=0$, the orbital energies in Eq.\ 
(\ref{ejm}) reduce to the corresponding $B=0$ result derived in
section III.A, namely $E_1=\epsilon-2\beta$ and 
$E_2=E_3=\epsilon+\beta$. Notice, however, that for arbitrary values of 
$B$, the degeneracy between $E_2$ and $E_3$ is lifted. In addition, the three
energies $E_1$, $E_2$, and $E_3$ in Eq.\ (\ref{ejm}) exhibit 
prominent Aharonov-Bohm (AB) oscillations. It is interesting to compare this 
behavior to the behavior of the calculated canonical UHF orbital energies for 
$R_W=10$. These UHF orbital energies as a function of $B$ are displayed in 
Fig.\ 7. An inspection of Fig.\ 7 reveals that the UHF orbital energies do
exhibit (as expected) an Aharonov-Bohm oscillation as a function of $B$. 
However, these oscillations are more complicated from what can be simply
anticipated from the analytic formulas in Eq.\ (\ref{ejm}). 
Namely, the amplitude of the UHF AB oscillations decreases as $B$ increases.
This behavior is due to a decrease in the hopping parameter $\beta$,
which results from the spatial shrinkage of the Gaussian-type UHF orbitals
as a function of $B$. Eventually, for $B \rightarrow \infty$, all three
energies are degenerate. We notice that complete degeneracy of all UHF 
orbitals for any $N$ appears also in the $B=0$, $R_W \rightarrow \infty$ limit.

The electronic structure of the UHF fully-polarized three-electron molecule in
a magnetic field, which was discussed above and which exhibits Aharonov-Bohm
oscillations, does not have an analog in the realm of natural molecules. 
However, apart from the $B$-dependence of $\beta$, it agrees in a remarkable 
way with the ``noninteracting spectra'' of the artificial molecules that
can be formed out of 1D ring arrays of single QD's.\cite{kot}

\subsection{The $S_z=1/2$ partially polarized case}

\begin{figure}[t]
\centering\includegraphics[width=8.5cm]{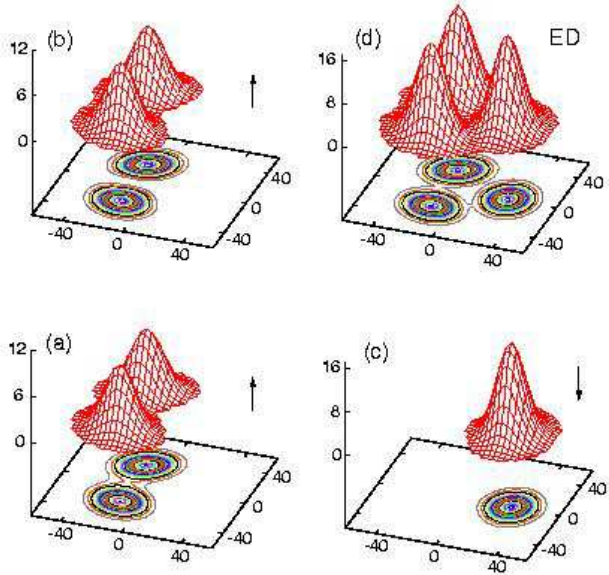}\\
~~~~\\
\caption{
The sS-UHF solution exhibiting breaking of the circular symmetry for 
$N=3$,$S_z=1/2$ at $R_W=10$ and $B=2$ T. (a-b): orbitals (modulus 
square) for the two spin-up electrons. (c): orbital (modulus square) for the 
spin-down electron. (d): total electron density.
The choice of the remaining parameters is: $\hbar \omega_0=5$ meV,
$m^*=0.067 m_e$, and $g^*=-0.44$.
Distances are in nanometers and the densities (orbital and ED) in 10$^{-4}$ 
nm$^{-2}$.
The arrows indicate the spin direction.
}
\end{figure}
Fig.\ 8 displays the BS UHF orbitals (modulus square) and the total ED for the 
partially polarized $N=3$, $S_z=1/2$ case in a magnetic field $B=2$ T. 
As with the $B=0$ case (section III.B), the spin-down orbital is decoupled
from the two spin-up ones. As a result the corresponding H\"{u}ckel matrix
equation is of the form
\begin{equation}
\left( \begin{array}{ccc}
             \epsilon  & -\beta e^{i\Omega}     & 0 \\
  -\beta e^{-i\Omega}  & \epsilon               & 0 \\
            0          &     0                  & \epsilon^\prime
       \end{array}   \right)
\left( \begin{array}{c}
          f_1 \\ f_2 \\ f_3
        \end{array}   \right)
= E \left( \begin{array}{c}
              f_1 \\ f_2 \\ f_3
           \end{array}   \right),
\label{meqn21m}
\end{equation}
where in general $\epsilon^\prime \neq \epsilon$ due to the energy difference 
between the two spin directions introduced by the Zeeman term.
From the solutions of Eq.\ (\ref{meqn21m}), one finds the
following LCAO-MO's: (a) $\psi_1 = (e^{i \Omega} \widetilde{\phi}_1 + 
\widetilde{\phi}_2)/\sqrt{2}$ with energy $E_1=\epsilon - \beta$, (b) 
$\psi_2 = (e^{i \Omega} \widetilde{\phi}_1 - \widetilde{\phi}_2)/\sqrt{2}$ 
with energy $E_2=\epsilon+\beta$, and (c) $\psi_3=\widetilde{\phi}_3$
with $E_3=\epsilon^\prime$. The total electron density constructed out of
these LCAO-MO's is again of the form $\phi_1^2+\phi_2^2+\phi_3^2$ 
(compare with section IV.A). The corresponding UHF orbitals (modulus square) 
and ED displayed in Fig.\ 8 are obviously conforming to these forms.

\begin{figure}[t]
\centering\includegraphics[width=7.5cm]{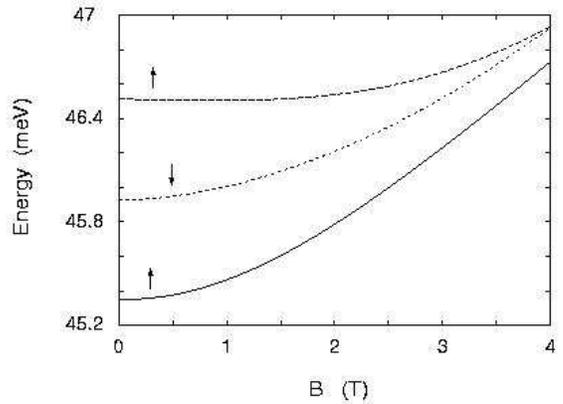}\\
~~~~\\
\caption{
The sS-UHF orbital energies (in meV) for $N=3$,$S_z=1/2$ and $R_W=10$ as a 
function of the magnetic field $B$ (in Tesla). No AB oscillations are present.
The choice of the remaining parameters is: $\hbar \omega_0 =5$ meV,
$m^*=0.067 m_e$, and $g^*=-0.44$. The arrows indicate the spin direction.
}
\end{figure}
Concerning the H\"{u}ckel orbital energies $E_j$, $j=1,2,3$, we note that 
they do not depend on the magnetic field $B$ through $\Omega$. As a result, 
unlike the previous case of the fully polarized electrons, AB oscillations 
should not develop in the UHF orbital energies. To check this prediction, we 
dispaly in Fig.\ 9 the UHF orbital energies as a function of $B$. In contrast
to Fig.\ 7, AB oscillations are absent in Fig.\ 9, a behavior which
apparently relates to the fact that no UHF orbital covers the area of the
equilateral triangle (the single spin-down orbital does not couple to the two 
spin-up ones, which lie on a straight line).

\section{Six electrons at zero magnetic field}

We discuss now the case of six fully polarized electrons in zero magnetic 
field. The corresponding total S-UHF electron density for $R_W=15$
$(\kappa=1.2730)$ is displayed
in Fig.\ 10(a) (bottom frame). Unlike the case of smaller numbers of particles 
with $N \leq 5$, six electrons is the smallest system that forms a Wigner 
molecule with a two-ring arrangement. Such a ring arrangement is denoted by 
$(1,5)$ to distinguish it from a single-ring arrangement $(0,N)$.

\begin{figure}[t]
\centering\includegraphics[width=8.5cm]{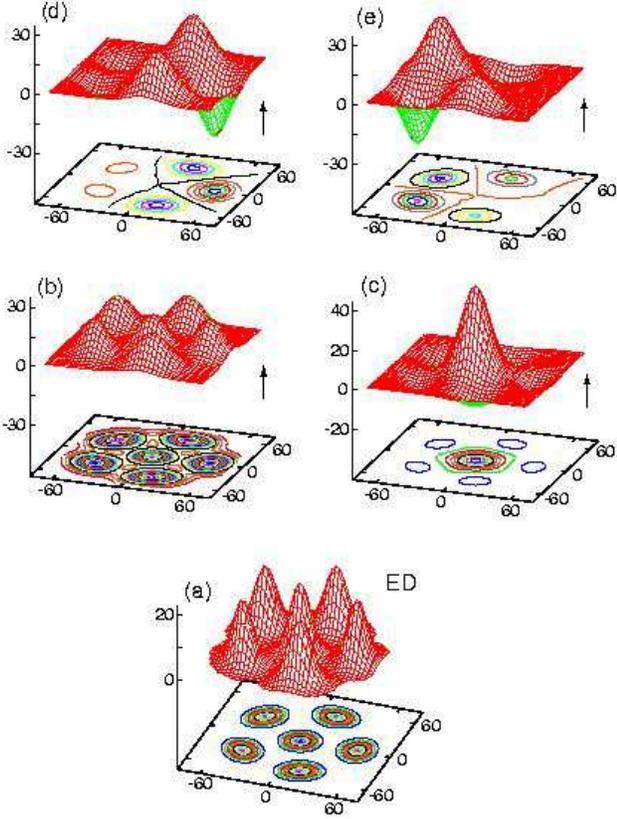}\\
~~~~\\
\caption{
The canonical S-UHF real orbitals for $N=6$ and $S_z=3$, 
and for $R_W=15$ and $B=0$.
(a): the total electron density. (b-c), middle row: the two orbitals
of $A$ symmetry. (d-e), top row: the two degenerate orbitals of $E_2$ symmetry.
The choice of the remaining parameters is: $\hbar \omega_0=5$ meV 
and $m^*=0.067 m_e$.
Distances are in nanometers. The real orbitals are in
10$^{-3}$ nm$^{-1}$ and the electron density in 10$^{-4}$ nm$^{-2}$.
The arrows indicate the spin direction.
}
\end{figure}
Naturally, single-ring molecular arrangements $(0,N)$ are familiar from the 
Quantum Chemistry of carbocyclic systems (Ref.\ \onlinecite{cot}, see also 
sections III and IV). The more complicated (1,5) arrangement, however, is a 
molecular structure unknown to traditional Chemistry. Nevertheless, contact to 
Organic Chemistry can be retained by observing that the (0,5) outer ring 
has a group symmetry similar to the cyclic hydrocarbon $C_5H_5$. As a result,
and in direct analogy with the $C_5H_5$ molecule (see p. 152 in Ref.\ 
\onlinecite{cot}), the SALC's of the (0,5) arrangement are as follows, 
\begin{equation}
\psi(A) = \frac{1}{\sqrt{5}}(\phi_1+\phi_2+\phi_3+\phi_4+\phi_5),
\label{phi61}
\end{equation}
\begin{eqnarray}
\psi(E_1a) = \sqrt{\frac{2}{5}}(&&\phi_1+\phi_2 \cos \theta
+\phi_3 \cos 2 \theta \nonumber \\
&& +\phi_4 \cos2 \theta +\phi_5 \cos \theta),
\label{phi62}
\end{eqnarray}
\begin{eqnarray}
\psi(E_1b) = \sqrt{\frac{2}{5}}(&&\phi_2 \sin \theta 
+\phi_3 \sin 2 \theta \nonumber \\
&& -\phi_4 \sin 2 \theta -\phi_5 \sin \theta),
\label{phi63}
\end{eqnarray}
\begin{eqnarray}
\psi(E_2a) = \sqrt{\frac{2}{5}}(&&\phi_1+\phi_2 \cos 2 \theta 
+\phi_3 \cos \theta \nonumber \\
&& +\phi_4 \cos  \theta +\phi_5 \cos 2 \theta),
\label{phi64}
\end{eqnarray}
\begin{eqnarray}
\psi(E_2b) = \sqrt{\frac{2}{5}}(&&\phi_2 \sin 2 \theta 
-\phi_3 \sin \theta \nonumber \\
&& +\phi_4 \sin  \theta -\phi_5 \sin 2 \theta), 
\label{phi65}
\end{eqnarray}
where $\theta = 2 \pi /5$.
The corresponding orbital energies are $\epsilon - 2 \beta $ for the 
single orbital of $A$ symmetry, $\epsilon - (2\cos \theta) \beta$ for
the two degenerate orbitals of $E_1$ symmetry, and $\epsilon - 
(2 \cos 2 \theta) \beta$ for the remaining two degenerate orbitals of $E_2$
symmetry. 

Returning back to the case of the (1,5) ring arrangement, we notice that
the sixth AO, $\phi_6$, at the center is of $A$ symmetry, and thus it can only
couple to a MO of the (0,5) ring with the same symmetry, namely the orbital 
$\psi(A)$ in Eq.\ (\ref{phi61}). As a result, both the (0,5) and the (1,5) 
ring arrangements share the same four MO's of $E_1$ and $E_2$ symmetry.

The coupling matrix element between the $\phi_6$ and $\psi(A)$ orbitals is 
given by, 
\begin{eqnarray}
\int \phi_6 \widetilde{H} \psi(A) d{\bf r} & =&  \frac{1}{\sqrt{5}}
\sum_{k=1}^5 \int \phi_6 \widetilde{H} \phi_k d{\bf r}  \nonumber \\
& = & - \sqrt{5} \delta.
\label{delt}
\end{eqnarray}

To find the MO's of the (1,5) ring with $A$ symmetry, we need to solve
the 2$\times$2 matrix equation,
\begin{equation}
\left( \begin{array}{cc}
         \epsilon - 2 \beta & -\sqrt{5} \delta \\
           -\sqrt{5} \delta & \widetilde{\epsilon}  
       \end{array}   \right)
\left( \begin{array}{c}
          g_1 \\ g_2 
        \end{array}   \right)
= E \left( \begin{array}{c}
              g_1 \\ g_2 
           \end{array}   \right).
\label{meqn665}
\end{equation}

We note that, due to the different coordination and distances, the quantities
$\delta$ and $\widetilde{\epsilon}$ associated with the central AO are
different from the corresponding quantities $\beta$ and $\epsilon$ 
associated with the AO's of the outer ring.

Using the notation,
\begin{equation}
Q = \sqrt{ 20 \delta^2 + (\widetilde{\epsilon}-\epsilon+2\beta)^2},
\label{qq}
\end{equation}
the two solutions of the matrix equation (\ref{meqn665}) have energies
$(\widetilde{\epsilon}+\epsilon-2\beta \mp Q)/2$ and eigenvectors 
(unnormalized) $\{\widetilde{\epsilon} - \epsilon + 2\beta \pm Q,
\sqrt{20} \delta \}$, respectively. 
Accordingly, one MO of the (1,5) ring is constructed by adding and the other 
by subtracting a fraction of the $\psi(A)$ orbital from the central AO. This 
behavior agrees very well with the two S-UHF orbitals displayed in the second
row of Fig.\ 10 [(b) and (c)]. 

Concerning the S-UHF orbitals displayed in the 
third row of Fig.\ 10 [(d) and (e)], we notice that they are degenerate in 
energy and that they agree very well with the two MO's of $E_2$ symmetry. 
Indeed the S-UHF orbital in Fig.\ 10(d) exhibits five total humps, three of 
them positive and the other two negative, in remarkable agreement (apart from 
an overall sign) with the MO in Eq.\ (\ref{phi64}) [note that $\cos 
\theta = 0.3090 > 0$ and $\cos 2\theta = -0.8090 <0$]. In particular, 
counterclockwise, the polarities of the humps in Fig.\ 10(d) are 
($-,+,-,-,+$), differing only by an overall sign from the corresponding 
polarities of the MO in Eq.\ (\ref{phi64}). Even more impressive is the
fact that there is quantitative agreement regarding the absolute heights of 
the humps in these two orbitals (see the values of $\cos \theta$ and $\cos
2 \theta$ listed above). 
The other S-UHF orbital in Fig.\ 10(e), exhibits a total of four humps,
two of them positive and the other two negative, and having an
alternating $(+,-,+,-)$ arrangement. This is again in remarkable agreement
with the second MO of $E_2$ symmetry in Eq.\ (\ref{phi65}), since 
$\sin \theta > 0$ and $\sin 2 \theta > 0$. Additionally, we note that the 
agreement between the UHF orbital in 
Fig.\ 10(e) and the MO in Eq.\ (\ref{phi65}) extends further to the absolute 
heights of the humps, since $\sin \theta =0.9511 > 0.5878 = \sin 2 \theta$.

Finally, there are two other degenerate UHF orbitals that are not displayed in
Fig.\ 10. They are not identical to the $\psi(E_1a)$ and $\psi(E_1b)$ SALC's
in Eqs.\ (\ref{phi62}) and (\ref{phi63}), but we have checked that they
span the $E_1$ irreducible representation. 

\section{Restoration of circular symmetry and exact spectra}

\subsection{Group structure and sequences of magic angular momenta}

In the previous sections, we demonstrated that the BS UHF determinants and 
orbitals describe indeed 2D electronic molecular stuctures (Wigner molecules) 
in close analogy with the case of natural 3D molecules. However, the study of 
the WM's at the UHF level restricts their description to the {\it intrinsic\/}
(nonrotating) frame of reference. Motivated by the case of natural atoms, one 
can take a subsequent step and address the properties of 
{\it collectively\/} rotating WM's in the 
laboratory frame of reference. As is well known, for natural atoms, this step 
is achieved by writing the total wave function of the molecule as the product
of the electronic and ionic partial wave functions. In the case of the purely
electronic WM's, however, such a product wave function requires the assumption
of complete decoupling between intrinsic and collective degrees of freedom,
an assumption that might be justifiable in limiting cases only.

As we demonstrated earlier,\cite{yl5,yl6,yl9} in the framework of the BS UHF 
solutions this companion step can be performed by using the 
post-Hartree-Fock method of {\it restoration of broken symmetries\/}\cite{rs} 
(RBS) via projection techniques (PT's). Examples demonstrating the RBS method 
have been presented by us in two cases: (I) The ground state (with angular
momentum $I=0$) of two interacting electrons in a parabolic quantum dot in the
absence of a magnetic field;\cite{yl5} (II) The yrast rotational band (see 
section I.C and precise definition in Ref.\ \onlinecite{note10}) 
of a system of $N$ interacting electrons in high magnetic fields\cite{yl6,yl9}
(fractional-quantum-Hall regime). In both cases, we showed 
that the RBS method (as adapted to the case of 2D BS UHF solutions) yields 
correlated (multideterminantal) many-body wave functions that approximate very
well the corresponding exact solutions.\cite{note10}
In particular, in the latter case, our use\cite{yl6,yl9} 
of the RBS method yielded analytic expressions for the 
correlated wave functions that offer a better description of the $N$-electron 
problem in high $B$ compared to the Jastrow-Laughlin\cite{lau2} expression.

In this section, we will not proceed any further with explicit numerical or 
analytic derivations of additional RBS wave functions. Instead, we will use 
the RBS approach to illustrate through a couple of concrete examples how 
certain universal properties of the exact solutions, i.e., the appearance of 
magic angular momenta in the exact rotational spectra, 
\cite{gir,mc,rua,sek,mak,haw} relate to the symmetry broken UHF solutions. 
Indeed, {\it we will demonstrate that the magic angular momenta are a direct 
consequence of the symmetry breaking at the UHF level and that they are 
determined fully by the molecular symmetries of the UHF 
determinant.\/}\cite{note12}  

As an illustrative example, we have chosen the relatively simple, but non 
trivial case, of $N=3$ electrons. For $B=0$, both the $S_z=1/2$ and $S_z=3/2$
polarizations can be considered. We start with the $S_z=1/2$ polarization,
whose BS UHF solution (let's denote it by $|\downarrow \uparrow \uparrow 
\rangle$) was presented in section III.B and which exhibits
a breaking of the total spin symmetry in addition to the rotational symmetry.
We first proceed with the restoration of the total spin by noticing that
$|\downarrow \uparrow \uparrow \rangle$ has a point-group symmetry lower 
(see section III.B) than the $C_{3v}$ symmetry of an equilateral triangle. 
The $C_{3v}$ symmetry, however, can be readily restored by applying the 
projection operator (\ref{prch}) to $|\downarrow \uparrow \uparrow \rangle$ 
and by using the character table of the cyclic $C_3$ group (see Table I). Then 
for the intrinsic part of the many-body wave function, one finds two different 
three-determinantal combinations, namely
\begin{equation}
\Phi_{\text{intr}}^{E^\prime} (\gamma_0)= 
|\downarrow \uparrow \uparrow \rangle
+ e^{2\pi i/3} |\uparrow \downarrow \uparrow \rangle
+ e^{-2\pi i/3} |\uparrow \uparrow \downarrow \rangle,
\label{3dete1}
\end{equation}
and
\begin{equation}
\Phi_{\text{intr}}^{E^{\prime\prime}} (\gamma_0)=
|\downarrow \uparrow \uparrow \rangle
+ e^{-2\pi i/3} |\uparrow \downarrow \uparrow \rangle
+ e^{2\pi i/3} |\uparrow \uparrow \downarrow \rangle,
\label{3dete2}
\end{equation}
where $\gamma_0=0$ denotes the azimuthal angle of the vertex associated with
the original spin-down orbital in $|\downarrow \uparrow \uparrow \rangle$.
We note that the intrinsic wave functions $\Phi_{\text{intr}}^{E^\prime}$
and $\Phi_{\text{intr}}^{E^{\prime\prime}}$ are eigenstates of the square of
the total spin operator ${\hat{\bf S}}^2$  ($\hat{\bf S} = \sum_{i=1}^3
\hat{\bf s}_i$) with quantum number $s=1/2$. This can be verified directly by 
applying ${\hat {\bf S}}^2$ to them.\cite{note8}

To restore the circular symmetry in the case of a (0,N) ring arrangement, one 
applies the projection operator,\cite{yl5,rs}
\begin{equation}
2 \pi {\cal P}_I \equiv \int_0^{2 \pi}
d\gamma \exp[-i \gamma (\hat{L}-I)]~,
\label{amp}
\end{equation}
where $\hat{L}=\sum_{j=1}^N \hat{l}_j$ is the operator for the total angular
momentum. Notice that the operator ${\cal P}_I$ is a direct generalization of
the projection operator (\ref{prch}) to the case of the continuous cyclic
group $C_\infty$ [the phases $\exp(i \gamma I)$ are the characters of 
$C_\infty$].

The RBS projected wave function, $\Psi_{\text{RBS}}$, (having both good total 
spin and angular momentum quantum numbers) is of the form,
\begin{equation}
2 \pi \Psi_{\text{RBS}} = \int^{2\pi}_0 d\gamma
\Phi_{\text{intr}}^E (\gamma) e^{i\gamma I},
\label{rbsi}
\end{equation}
where now the intrinsic wave function [given by Eq.\ (\ref{3dete1}) or
Eq.\ \ref{3dete2})] has an arbitrary azimuthal orientation $\gamma$. We note 
that, unlike the phenomenological Eckardt-frame model\cite{mak,note2} where 
only a single product term is involved, the RBS wave function in Eq.\ 
(\ref{rbsi}) is an average over all azimuthal directions of an infinite
set of product terms. These terms are formed by multiplying the UHF intrinsic 
part $\Phi_{\text{intr}}^{E}(\gamma)$ with the external rotational wave 
function  $\exp(i \gamma I)$ (the latter is properly
characterized as ``external'', since it is an eigenfunction of the total
angular momentum $\hat{L}$ and depends exclusively on the azimuthal
coordinate $\gamma$).

The operator ${\hat{R}(2\pi/3) \equiv \exp (-i 2\pi{\hat L}/3})$ can be
applied onto $\Psi_{\text{RBS}}$ in two different ways, namely either on
the intrinsic part $\Phi_{\text{intr}}^{E}$ or the external part $\exp(i \gamma
I)$. Using Eq.\ (\ref{3dete1}) and the property $\hat{R}(2\pi/3) 
\Phi_{\text{intr}}^{E^\prime} =\exp (-2\pi i/3)\Phi_{\text{intr}}^{E^\prime}$, 
one finds,
\begin{equation}
\hat{R}(2\pi/3) \Psi_{\text{RBS}} = \exp (-2\pi i/3) \Psi_{\text{RBS}},
\label{r1rbs}
\end{equation}
from the first alternative, and
\begin{equation}
\hat{R}(2\pi/3) \Psi_{\text{RBS}} = \exp (-2\pi I i/3) \Psi_{\text{RBS}},
\label{r2rbs}
\end{equation}
from the second alternative. Now if $\Psi_{\text{RBS}} \neq 0$, the only
way that Eqs.\ (\ref{r1rbs}) and (\ref{r2rbs}) can be simultaneously true is
if the condition $\exp [2\pi (I-1) i/3]=1$ is fulfilled. This leads to a first
sequence of magic angular momenta associated with total spin $s=1/2$, i.e.,
\begin{equation}
I = 3 k +1,\; k=0,\pm 1, \pm 2, \pm 3,...
\label{i1}
\end{equation}

Using Eq.\ (\ref{3dete2}) for the intrinsic wave function, and following
similar steps, one can derive a second sequence of magic angular momenta
associated with good total spin $s=1/2$, i.e.,
\begin{equation}
I = 3 k -1,\; k=0,\pm 1, \pm 2, \pm 3,...
\label{i2}
\end{equation}

In the fully polarized case, the UHF determinant was described in section 
III.A. This UHF determinant, which we denote as $|\uparrow \uparrow \uparrow
\;\rangle$, is already an eigenstate of $\hat{\bf S}^2$ with quantum number 
$s=3/2$. Thus only the rotational symmetry needs to be restored, that is,
the intrinsic wave function is simply $\Phi^A_{\text{intr}}(\gamma_0) = 
|\uparrow \uparrow \uparrow \;\rangle$. Since 
$\hat{R}(2\pi/3) \Phi^A_{\text{intr}} = \Phi^A_{\text{intr}}$, the condition
for the allowed angular momenta is $\exp [-2\pi I i/3]=1$, which yields
the following magic angular momenta,
\begin{equation}
I = 3 k,\; k=0,\pm 1, \pm 2, \pm 3,...
\label{i3}
\end{equation}

We note that in high magnetic fields only the fully polarized case is
relevant and that only angular momenta with $k > 0$ enter in Eq.\ (\ref{i3})
(see Ref.\ \onlinecite{yl6}). In this case, in the thermodynamic limit, the 
partial sequence with $k=2q+1$, $q=0,1,2,3,...$ is directly related to the odd 
filling factors $\nu=1/(2q+1)$ of the fractional quantum Hall effect
[via the relation $\nu = N(N-1)/(2I)$]. This suggests that the observed 
hierarchy of fractional filling factors in the quantum Hall effect may be 
viewed as a signature originating from the point group symmetries of the 
intrinsic wave function $\Phi_{\text{intr}}$, and thus it is a manifestation
of symmetry breaking at the UHF mean-field level. 

\begin{table}[t]
\caption{%
Case of $N=6$ electrons in high magnetic field $B$: Total interaction energies
in the lowest Landau level of REM and exact-diagonalization wave functions for
various magic angular momenta $I$ of the yrast band. The REM functions are 
analytically specified RBS wave functions derived in Ref.\ 
\protect\onlinecite{yl6}. The percentages within parenthesis indicate 
relative errors. Energies in units of $e^2/\kappa l_B$, where $\kappa$ is
the dielectric constant and $l_B=\sqrt{\hbar c/eB}$ is the magnetic length.
For details concerning the exact-diagonalization method and the REM wave
functions, see Ref.\ \protect\onlinecite{yl9}. For additional values of $I$,
see Ref.\ \protect\onlinecite{yl9}.%
}
\begin{tabular}{lcr}
$I$ & REM & EXACT \\ \hline
70  & 2.3019 (0.85\%) & 2.2824 \\
80  & 2.1455 (0.71\%) & 2.1304 \\
90  & 2.0174 (0.60\%) & 2.0053 \\
100 & 1.9098 (0.51\%) & 1.9001 \\
110 & 1.8179 (0.45\%) & 1.8098 \\
120 & 1.7382 (0.40\%) & 1.7312 \\
130 & 1.6681 (0.36\%) & 1.6621 \\
\end{tabular}
\end{table}

\subsection{Quantitative description of the yrast band}

The usefulness of the RBS wave functions [Eq.\ (\ref{rbsi})] is not limited to
deriving universal properties of the exact spectra, like the sequences of 
magic angular momenta [see section VI.A]. As we demonstrated in earlier 
publications, in the the regime of strong correlations, the RBS wave functions
approximate very well the corresponding exact many-body wave functions. 

Indeed, in Ref.\ \onlinecite{yl5} we offered (as a function of $R_W$) a 
systematic comparison between the RBS and exact ground-state ($I=0$) energies 
at $B=0$ for $N=2$ electrons in a parabolic QD. For $R_W=19.09$, 
we found that the relative error was approximately 0.7\%. Furthermore
in Ref.\ \onlinecite{yl6}, for the case of high $B$, we derived 
{\it analytic\/} RBS wave functions, named ``REM wave functions''. 
As we showed\cite{yl6} explicitly for the case of $N=6$ electrons, the radial 
electron densities associated with the REM functions accurately reproduce the 
ones extracted from exact-diagonalization calculations.

In this subsection, we offer additional examples pertaining to the ability of 
the RBS wave functions to reproduce the exact yrast spectra of parabolic
QD's. In particular, Table III lists the REM and exact yrast energies in
the range of magic angular momenta $70 \leq I \leq 130$ for $N=6$ 
electrons in high $B$. Details concerning the REM wave functions and the
exact-diagonalization method in the lowest Landau level are given in Ref.\
\onlinecite{yl9}, and they will not be repeated here. 

The RBS and exact yrast spectra ($0 \leq I \leq 6$) for the case of $N=2$ 
electrons at $B=0$ and $R_W=19.09$ are given in Table IV. Details concerning 
their calculation are given in Ref.\ \onlinecite{yl5} and in the Appendix 
(where we present the final formula for calculating RBS energies for both even
and odd angular momenta).

We note that the relative errors in both Table III and Table IV are small
(smaller than 1\% in the majority of cases).

\section{Summary}

\begin{table}[t]
\caption{%
Case of $N=2$ electrons in a parabolic QD at $B=0$ : Total energies of RBS and 
exact wave functions for various magic angular momenta $I$ of the yrast band.
The percentages within parenthesis indicate relative errors. 
The choice of remaining parameters is: $\hbar \omega_0=5$ meV, $\kappa=1$
($R_W =19.09$), $m^*=0.067 m_e$. Energies in units of meV. For details 
concerning the method for finding exact solutions to the two-electron problem,
see Ref.\ \protect\onlinecite{yl7}. For details concerning
the calculation of the RBS yrast spectrum, see Ref.\ 
\protect\onlinecite{yl5} and the Appendix.
}
\begin{tabular}{lcr}
$I$ & RBS & EXACT \\ \hline
0 & 52.224 (0.75\%) & 51.831 \\
1 & 52.696 (0.77\%) & 52.292 \\
2 & 54.086 (0.88\%) & 53.615 \\
3 & 56.240 (1.04\%) & 55.654 \\
4 & 59.065 (1.39\%) & 58.255 \\
5 & 62.065 (1.27\%) & 61.285 \\
6 & 63.911 (1.13\%) & 64.642 \\
\end{tabular}
\end{table}

In this paper, we have introduced a group theoretical analysis of 
broken-symmetry UHF orbitals and total electron densities in the case of 
single 2D semiconductor QD's. This analysis provided further support for our 
earlier interpretation\cite{yl3,yl4,yl7,yl5} concerning the spontaneous 
formation of collectively rotating electron (or Wigner) molecules. Indeed the 
group-theoretical analysis enabled us to unveil further deeper analogies 
between the electronic structure of the Wigner molecules and that of the 
natural 3D molecules. In particular these deeper analogies are: 
(I) The breaking of rotational 
symmetry results in canonical UHF orbitals that are associated with the 
eigenvectors of a molecular-type H\"{u}ckel hamiltonian with sites at
positions specified by the equilibrium configuration of the classical 
$N$-electron problem; (II) The broken-symmetry canonical UHF orbitals 
transform according 
to the irreducible representations of the point group specified by the 
discrete symmetries of this classical molecular configuration;
(III) The WM's formed out of the broken-symmetry UHF solutions can 
rotate, and the restoration of the total-spin and rotational symmetries 
results (in addition to the ground state) in states defining the lowest 
rotational bands (i.e., yrast bands) of the WM's; (IV) The breaking of
the circular symmetry results in lowering of the symmetry. This is expressed
by the discrete point-group symmetry of the UHF wave function and it underlies
the appearance of sequences of magic angular momenta (familiar from
exact-diagonalization studies) in the excitation spectra of single QD's. 

Since exact-diagonalization methods are typically restricted to small sizes
with $N \leq 10$, the two-step method of breakage and subsequent
restoration of symmetries offers a promising new venue for accurately
describing larger 2D electronic systems. A concrete example of the potential
of this approach is provided by Ref.\ \onlinecite{yl6}, where
our use of the the symmetry-breaking/symmetry-restoration method yielded
analytic expressions for correlated wave functions that offer a better
description of the $N$-electron problem in high magnetic fields compared
to the Jastrow-Laughlin\cite{lau2} expression.

Furthermore, the group-theoretical analysis strongly suggests an interesting 
simplified variant approach for carrying out the first step of symmetry 
breaking. This variant rests on the observation that, in all cases of WM's, 
the broken-symmetry UHF orbitals are generic linear combinations of 
Gaussian-type functions [with a proper phase for $B \neq 0$, see Eq.\ 
(\ref{phase})] specified simply by their width $\sigma$ and positions 
${\bf R}_j$'s from the center of the QD. The linear combinations can be fully 
specified from the group theoretical analysis of the appropriate classical
equilibrium configurations,\cite{bp} and a determinant of the 
corresponding LCAO-MO's can readily be written down. Then a simple variational
calculation of the minimum total energy of this determinant will yield the 
parameters $\sigma$ and ${\bf R}_j$'s without the need to carry out the 
self-consistent UHF iterations. This simplified approach could treat even 
larger sizes without major loss of accuracy. Added accuracy can then be 
obtained through the subsequent step of restoration of the broken symmetries.

\acknowledgments
This research was supported by a grant from the U.S. Department of Energy
(Grant No. FG05-86ER-45234).

\appendix
\section{}

For the case of $N=2$ electrons in a parabolic QD at $B=0$, we reported in 
Ref.\ \onlinecite{yl5} the RBS formulas for calculating energies of 
yrast-band states with {\it even\/} angular momenta $I$. These formulas [see
Eqs. (11)-(13) in Ref.\ \onlinecite{yl5}] were generated via a projection of 
the ``singlet'' UHF determinant. The corresponding RBS formulas for 
{\it odd\/} values of $I$ are generated via a projection of the triplet UHF 
state. 

In this Appendix, we present the formulas covering both even and odd angular
momenta. They are:

\begin{equation}
E_{\text{RBS}} (I) = \left. { \int_0^{2\pi} h(\gamma) e^{i \gamma I} 
d\gamma } \right/ { \int_0^{2\pi} n(\gamma) e^{i \gamma I} d\gamma},
\label{eproj}
\end{equation}
with
\begin{eqnarray}
h(\gamma) =&&
H_{us}S_{vt} \pm H_{ut}S_{vs} + H_{vt}S_{us} \pm H_{vs}S_{ut}+ \nonumber \\
 && V_{uvst} \pm V_{uvts},
\label{hgam}
\end{eqnarray}
and
\begin{equation}
n(\gamma)= S_{us}S_{vt} \pm S_{ut}S_{vs},
\label{ngam}
\end{equation}
where the upper signs apply in the case of even $I$'s and the lower signs 
in the case of odd $I$'s. 
$s({\bf r})$ and $t({\bf r})$ are the initial $u({\bf r})$ and $v({\bf r})$
broken-symmetry UHF orbitals rotated by an angle $\gamma$, respectively.
$V_{uvst}$ and $V_{uvts}$ are two-body matrix elements of the Coulomb
repulsion, and $S_{us}$, etc., are the overlap intergrals.


\begin{references}
\bibitem{kas}
M.A. Kastner,
Phys. Today {\bf 46(1)}, 24 (1993).
\bibitem{ash}
R.C. Ashoori,
Nature {\bf 379}, 413 (1996).
\bibitem{tar}
S. Tarucha, D.G. Austing, T. Honda, R.J. van der Hage, and L.P. Kouwenhoven,
Phys. Rev. Lett. {\bf 77}, 3613 (1996).
\bibitem{kou}
L.P. Kouwenhoven, C.M. Marcus, P.L. McEuen, S. Tarucha, 
R.M. Westervelt, and N.S. Wingreen, 
Proceedings of the NATO Advanced Study Institute on {\it Mesoscopic Electron 
Transport\/}, Series E, Vol. 345, edited by L.L. Sohn, L.P. Kouwenhoven, 
and G. Sch\"{o}n (Kluwer, Dordrecht, 1997) p. 105.
\bibitem{lau2}
R.B. Laughlin,
Phys. Rev. Lett. {\bf 50}, 1395 (1983).
\bibitem{cs}
E.U. Condon and G.H. Shortley,
{\it The theory of atomic spectra\/}
(Cambridge Univ. Press, London, 1935).
\bibitem{hart}
D.R. Hartree,
Proc. Camb. Phil. Soc. {\bf 24}, 89 (1928).
\bibitem{thtr}
See, e.g., exact calculations by M. Eto, Jpn.
J. Appl. Phys. {\bf 36}, 3924 (1997); Hartree-Fock calculations
by A. Natori, Y. Sugimoto, and M. Fujito, Jpn. J. Appl. Phys. {\bf 36}, 3960
(1997); H. Tamura, Physica (Amsterdam) {\bf 249B-251B}, 210
(1998); and M. Rontani, F. Rossi, F. Manghi, and E. Molinari,
Phys. Rev. B {\bf 59}, 10 165
(1999); the spin density-functional theory at $B=0$ by
In-Ho Lee, V. Rao, R.M. Martin, and J.P. Leburton,
Phys. Rev. B {\bf 57}, 9035 (1998) and Ref.\ \onlinecite{kmr};
and also at nonzero $B$ by O. Steffen,
U. R\"{o}ssler, and M. Suhrke, Europhys. Lett. {\bf 42}, 529
(1998).
\bibitem{yl3}
C. Yannouleas and U. Landman,
Phys. Rev. Lett. {\bf 82}, 5325 (1999).
\bibitem{yl4}
C. Yannouleas and U. Landman,
Phys. Rev. B {\bf 61}, 15895 (2000).
\bibitem{bp}
For studies pertaining to the geometrical arrangements of classical 
point charges in a harmonic confinement, see Yu.E. Lozovik,
Usp. Fiz. Nauk {\bf 153}, 356 (1987)
[Sov. Phys. Usp. {\bf 30}, 912 (1987)] and
V.M. Bedanov and F.M. Peeters,
Phys. Rev. B {\bf 49}, 2667 (1994).
\bibitem{hg}
For parabolic QD's at zero magnetic field with $6 \leq N \leq 8$, a crossover 
from singly-humped to doubly-humped radial electron densities was found as a
function of $R_W$ via path-integral Monte Carlo simulations [R. Egger, 
W. H\"{a}usler, C.H. Mak, and H. Grabert, Phys. Rev. Lett. {\bf 82}, 3320 
(1999)]. Since the doubly-humped radial electron densities are compatible with
the $(1,N-1)$ polygonal structures of classical point charges in the range  
$6 \leq N \leq 8$ (after carrying out an azimuthal averaging), this crossover 
was interpreted as indicating formation of WM's.
\bibitem{loz}
These transitions were further elaborated in 
A. V. Filinov, M. Bonitz, and Yu. E. Lozovik, 
Phys. Rev. Lett. 86, 3851 (2001).
\bibitem{yl7}
C. Yannouleas and U. Landman,
Phys. Rev. Lett. {\bf 85}, 1726 (2000).
\bibitem{yl5}
C. Yannouleas and U. Landman,
J. Phys.: Condens. Matter {\bf 14}, L591 (2002).
\bibitem{yl6}
C. Yannouleas and U. Landman,
Phys. Rev. B {\bf 66}, 115315 (2002).
\bibitem{yl9}
C. Yannouleas and U. Landman,
arXiv.org: cond-mat/0302504.
\bibitem{note7}
Unlike the HF approach for which a fully developed theory for the restoration
of symmetries has long been established (see section I.B), the breaking of 
symmetries within the spin-dependent density functional theory poses a
serious dilemma 
[J.P. Perdew, A. Savin, and K. Burke, Phys. Rev. A {\bf 51}, 4531 (1995)].
This dilemma has not been resolved todate; several remedies
(like Projection, ensembles, etc.) are being proposed, but none of them
appears to be completely devoid of inconsistencies [A. Savin, in
{\it Recent Developments and Applications of Modern Density Functional
Theory\/}, edited by J.M. Seminario (Elsevier, Amsterdam, 1996), p. 327].
In addition, due to the unphysical self interaction energy (which vigorously 
and {\it erroneously\/} assists the kinetic energy in orbital delocalization), 
the density-functional theory is more resistant against space symmetry breaking
[R. Bauernschmitt and R. Ahlrichs, J. Chem. Phys. {\bf 104}, 9047 (1996)]
than the sS-UHF, and thus it fails to describe a whole class of broken 
symmetries involving electron localization, e.g., the formation at $B=0$ of 
Wigner molecules in QD's (see footnote 7 in Ref.\ \onlinecite{yl3}),
the hole trapping at Al impurities in silica [J. Laegsgaard and K. Stokbro,
Phys. Rev. Lett. {\bf 86}, 2834 (2001); G. Pacchioni, F. Frigoli, D. Ricci, 
and J.A. Weil, Phys. Rev. B {\bf 63}, 054102 (2001)], or the interaction driven
localization-delocalization transition in $d$- and $f$- electron systems
[see, e.g., {\it Strong Coulomb Correlations in Electronic Structure 
Calculations : Beyond the Local Density Approximation\/}, edited by V.I. 
Anisimov (Gordon \& Breach, Amsterdam, 2000); S.Y. Savrasov, G. Kotliar, and E.
Abrahams, Nature {\bf 410}, 793 (2001)]. In line with the above, no density 
functional calculations describing space symmetry breaking and formation of 
Wigner molecules at $B=0$ in {\it circular\/} QD's have been reported todate.
\bibitem{bm1}
\AA. Bohr and B.R. Mottelson,
K. Danske Vidensk. Selsk. Mat.-Fys. Medd. {\bf 27}, no. 16 (1953).
\bibitem{nil}
S.G. Nilsson,
K. Danske Vidensk. Selsk. Mat.-Fys. Medd. {\bf 29}, no. 16 (1955).
\bibitem{cle}
K.L. Clemenger,
Phys. Rev. B {\bf 32}, 1359 (1985).
\bibitem{str}
V.M. Strutinsky,
Nucl. Phys. {\bf A95}, 420 (1967); Nucl. Phys. {\bf A122}, 1 (1968).
\bibitem{yl1}
C. Yannouleas and U. Landman,
Phys. Rev. B {\bf 51}, 1902 (1995).
\bibitem{yl2}
C. Yannouleas, U. Landman, and R.N. Barnett,
in {\it Metal Clusters\/}, edited by W. Ekardt (Wiley, Chichester, 1999)
p. 145.
\bibitem{rs}
P. Ring and P. Schuck, {\it The Nuclear Many-body Problem\/}
(Springer-Verlag, New York, 1980). 
\bibitem{so}
A. Szabo and N.S. Ostlund
{\it Modern Quantum Chemistry\/}
(McGraw-Hill, New York, 1989).
\bibitem{rs2}
See in particular Ch. 5.5 and Ch. 11 in Ref.\ \onlinecite{rs}.
However, our terminology (i.e., UHF vs. RHF) follows the practice in
Quantum Chemistry (see Ref.\ \onlinecite{so}).
\bibitem{mk}
For the case of high-magnetic fields, see also 
H.-M. M{\" u}ller and S.E. Koonin,
Phys. Rev. B {\bf 54}, 14532 (1996).
\bibitem{py}
For the restoration of broken rotational symmetries in atomic nuclei, see 
R.E. Peierls and J. Yoccoz, 
Proc. Phys. Soc. (London) {\bf A70}, 381 (1957), 
and Ch. 11 in Ref.\ \onlinecite{rs}.
\bibitem{low}
For the restoration of broken spin symmetries in natural 3D molecules, see
P.O. L\"{o}wdin,
Phys. Rev. B {\bf 97}, 1509 (1955);
Rev. Mod. Phys. {\bf 36}, 966 (1964).
\bibitem{fuk}
H. Fukutome,
Int. J. Quantum Chem. {\bf 20}, 955 (1981),
and references therein.
\bibitem{yl8}
For the restoration of broken spin symmetries in the case of
double QD's, leading to a Generalized Heitler-London approach 
for the coupling and dissociation of artificial molecules, see (for $B=0$) 
C. Yannouleas and U. Landman, Eur. Phys. J. D {\bf 16}, 373 (2001) and
(as a function of $B$) C. Yannouleas and U. Landman, Int. J. Quantum Chem. 
{\bf 90}, 699 (2002).
\bibitem{cot}
F.A. Cotton,
{\it Chemical Applications of Group Theory\/}
(Wiley, New York, 1990).
\bibitem{note9}
The use of the term yrast is customary in the spectroscopy
of rotating nuclei, see, e.g., 
\AA. Bohr and B.R. Mottelson,
{\it Nuclear Structure\/} (Benjamin, Reading, MA, 1975),
Vol. II, p. 41.
\bibitem{gir}
S.M. Girvin and T. Jach,
Phys. Rev. B {\bf 28}, 4506 (1983).
\bibitem{mc}
P.A. Maksym and T. Chakraborty,
Phys. Rev. Lett. {\bf 65}, 108 (1990).
\bibitem{rua}
W.Y. Ruan, Y.Y. Liu, C.G. Bao, and Z.Q. Zhang,
Phys. Rev. B {\bf 51}, 7942 (1995).
\bibitem{sek}
T. Seki, Y. Kuramoto, and T. Nishino,
J. Phys. Soc. Jpn. {\bf 65}, 3945 (1996).
\bibitem{mak}
P.A. Maksym, 
Phys. Rev. B {\bf 53}, 10871 (1996).
\bibitem{haw}
L. Jacak, P. Hawrylak, and A. Wojs,
{\it Quantum Dots\/} (Springer, Berlin, 1998), and references therein.
\bibitem{pn}
J.A. Pople and R.K. Nesbet,
J. Chem. Phys. {\bf 22}, 571 (1954).
\bibitem{note11}
The UHF equations preserve at each iteration step the symmetries
of the many-body hamiltonian, if these symmetries happen to
be present in the input (initial) electron density of the iteration (see 
section 5.5 of Ref.\ \onlinecite{rs}). The input densities into the iteration
cycle are controlled by the values of the $P^\alpha_{\lambda\sigma}$ and 
$P^\beta_{\lambda\sigma}$ matrix elements. (a) Symmetry adapted RHF solutions 
are extracted from Eq.\ (\ref{uhfa}) and Eq.\ (\ref{uhfb}) by using as input
$P^\alpha_{\lambda\sigma}=P^\beta_{\lambda\sigma}$=0 for the case of closed
shells (with or without an infinitesimally small $B$ value). For open shells,
one needs to use an infinitesimally small value of $B$. With these choices,
the output of the first iteration (for either closed or open shells) is the
single-particle spectrum and corresponding electron densities at $B=0$ 
associated with the hamiltonian in Eq.\ (\ref{hsp}) (the small value of $B$ 
mentioned above guarantees that the single-particle total and orbital 
densities are circular). (b) For obtaining broken-symmetry UHF solutions, the 
input densities must be different in an essential way from the ones mentioned 
above. We have found that the choice $P^\alpha_{\lambda\sigma}=1$ and 
$P^\beta_{\lambda\sigma}=0$ usually produces broken-symmetry solutions (in the
regime where symmetry breaking occurs).
\bibitem{kmr}
M. Koskinen, M. Manninen, and S.M. Reimann,
Phys. Rev. Lett. {\bf 79}, 1389 (1997).
\bibitem{note4}
In some circumstances, SDW's may be obtained from spin density functional
calculations (see Ref.\ \onlinecite{kmr}). In general, however, the
breakings of spin and/or spatial symmetries are not properly described within
spin density functional theory (see Ref.\ \onlinecite{note7}).
\bibitem{over}
The possibility of ground-state configurations with uniform electron density,
but nonuniform spin density, was first discussed for 3D bulk metals using
the HF method in A.W. Overhauser, Phys. Rev. Lett. {\bf 4}, 462 (1960); Phys. 
Rev. {\bf 128}, 1437 (1962).
\bibitem{note3}
Depending on the spin polarization, a WM may (or may not) be accompanied by 
a SDW. Unlike the pure SDW case, however, the SDW of a WM exhibits necessarily
the same number of humps as the number of electrons (see, e.g., the case of 
$N=3$ electrons in section III.B).
\bibitem{note5}
However, for $R_W \lesssim 1$, formation of a special class of SDW's (often 
called electron puddles) plays an important role in the coupling and 
dissociation of quantum dot molecules [see Ref.\ \onlinecite{yl8} and 
Ref.\ \onlinecite{yl3}].
\bibitem{note1}
In Solid State physics the H\"{u}ckel approximation is usually referred to as 
the tight-binding approximation, with $\beta$ denoted most often as $t$
[$t$ specifies the tunneling (hopping) between sites].
\bibitem{wol}
A.B. Wolbarst,
{\it Symmetry and Quantum Systems\/}
(Van Nostrand Reinold, New York, 1977).
\bibitem{note6}
The group theoretical symbol $E$ for the two-dimensional irreducible
representations should not be confused with the same symbol denoting the
eigenvalues of the H\"{u}ckel equation or the UHF orbital energies.
For the group theoretical symbols, we follow the Sch\"{o}nflies 
convention. In addition to $E$, for one-dimensional irreducible  
representations, we use the capital letters $A$ and $B$ [see Refs.\ 
\onlinecite{cot,wol}].
\bibitem{sun}
P.A. Sundqvist, S. Yu. Volkov, Yu. E. Lozovik, and M. Willander,
Phys. Rev. B {\bf 66}, 075335 (2002).
\bibitem{and}
S. Hammes-Schiffer and H.C. Andersen,
J. Chem. Phys. {\bf 99}, 1901 (1993).
\bibitem{bro}
E. Brown,
Phys. Rev. {\bf 133}, A1038 (1964).
\bibitem{zak}
J. Zak,
Phys. Rev. {\bf 134}, A1602 (1964).  
\bibitem{pei}
R.E. Peierls,
Z. Phys. {\bf 80}, 763 (1933).
\bibitem{har}
P.G. Harper, 
J. Phys.: Condens. Matter {\bf 3}, 3047 (1991).
\bibitem{kot}
R. Kotlyar, C.A. Stafford, and S. Das Sarma,
Phys. Rev. B {\bf 58}, 3989 (1998).
\bibitem{note10}
The RBS wave functions in Eq. (\ref{rbsi}) work for the group of 
states that exhibit magic angular momenta and have the lowest possible energy.
In our use of the term, the group of these states forms the
``yrast'' rotational band, namely the band of states whose excitation energy
represents {\it pure\/} rotational motion (no other excitations, like 
center-of-mass motion or vibrational modes are present).
\bibitem{note12}
We note that the {\it discrete\/} rotational (and more generally 
rovibrational) collective spectra associated with symmetry-breaking in a QD 
may be viewed as finite analogs to the Goldstone modes accompanying symmetry 
breaking transitions in extended media [see P.W. Anderson, {\it Basic Notions 
of Condensed Matter Physics\/} (Addison-Wesley, Reading, MA, 1984)].
\bibitem{note8}
For the appropriate expression of ${\bf S}^2$, see Eq.\ (6) of
C. Yannouleas and U. Landman, Int. J. Quantum Chem. {\bf 90}, 699 (2002).
\bibitem{note2}
Although the wave functions of the Eckardt-frame model are inaccurate compared
to the RBS ones [see Eq.\ (\ref{rbsi})], they are able to yield the proper 
magic angular momenta for $(0,N)$ rings. This result, however, is intuitively 
built in this model from the very beginning via the phenomenological 
assumption that the intrinsic wave function, which is never specified, 
exhibits $C_{Nv}$ point-group symmetries.

\end{references}
\end{document}